\documentclass[fleqn,usenatbib,useAMS]{mnras}

\usepackage{graphicx}	
\usepackage{amsmath}	
\usepackage{amssymb}	
\usepackage{multicol}        
\usepackage{bm}		
\usepackage{pdflscape}	
\usepackage{xcolor}     

\newcommand{\kms}{\,km\,s$^{-1}$} 
\newcommand{\vrad}{$v_\mathrm{rad}$}
\newcommand{\vsini}{$v_\mathrm{rot}\sin i$}
\newcommand{\msun}{$M_{\odot}$}

\newcommand{\teff}{$T_\mathrm{eff}$}
\newcommand{\logg}{$\log g$}
\newcommand{\logy}{$\log n(\mathrm{He})/n(\mathrm{H})$}

\newcommand{\jofour}{J0415+2538}
\newcommand{\jthirteen}{J1303+2646}
\newcommand{\jsixteen}{J1603+3412}
\newcommand{\joeight}{J0809-2627}
\newcommand{\tlusty}{\textsc{Tlusty}}

\newcommand{\uh}[1]{\textcolor{blue}{#1}}

\usepackage[T1]{fontenc}
\usepackage{ae,aecompl}

\usepackage{newtxtext,newtxmath}

\title[Discovery of three magnetic hot subdwarfs]{Discovery and analysis of three magnetic hot subdwarf stars: evidence for merger-induced magnetic fields}

\author[Pelisoli et al.]{Ingrid Pelisoli$^{1}$\thanks{E-mail: ingrid.pelisoli@warwick.ac.uk}, M. Dorsch$^{2}$, U. Heber$^{2}$, B. G\"{a}nsicke$^{1}$, S. Geier$^{3}$, T. Kupfer$^{4}$, P. N\'{e}meth$^{5,6}$, 
\newauthor S. Scaringi$^{7}$, V. Schaffenroth$^{3}$
\\
$^{1}$Department of Physics, University of Warwick, Gibbet Hill Road, Coventry, CV4 7AL, UK\\
$^{2}$Dr.\ Karl Remeis-Observatory \& ECAP, Friedrich-Alexander University Erlangen-N\"{u}rnberg,
Sternwartstr.\ 7, 96049 Bamberg, Germany\\
$^{3}$Institut für Physik und Astronomie, Universität Potsdam, Haus 28, $^{4}$Karl-Liebknecht-Str. 24/25, 14476 Potsdam-Golm, Germany\\
$^{4}$Texas Tech University, Department of Physics \& Astronomy, Box 41051, 79409, Lubbock, TX, USA\\
$^{5}$Astronomical Institute of the Czech Academy of Sciences, CZ-25165, Ond\v{r}ejov, Czech Republic\\
$^{6}$Astroserver.org, F\H{o} t\'er 1, 8533 Malomsok, Hungary\\
$^{7}$Centre for Extragalactic Astronomy, Department of Physics, Durham University, South Road, Durham, DH1 3LE, UK
}
\date{Last updated XXX; in original form XX}

\pubyear{2022}

\begin{document}
\label{firstpage}
\pagerange{\pageref{firstpage}--\pageref{lastpage}}
\maketitle

\begin{abstract}
Magnetic fields can play an important role in stellar evolution. Among white dwarfs, the most common stellar remnant, the fraction of magnetic systems is more than 20 per cent. The origin of magnetic fields in white dwarfs, which show strengths ranging from 40~kG to hundreds of MG, is still a topic of debate. In contrast, only one magnetic hot subdwarf star has been identified out of thousands of known systems. Hot subdwarfs are formed from binary interaction, a process often associated with the generation of magnetic fields, and will evolve to become white dwarfs, which makes the lack of detected magnetic hot subdwarfs a puzzling phenomenon. Here we report the discovery of three new magnetic hot subdwarfs with field strengths in the range 300--500~kG. Like the only previously known system, they are all helium-rich O-type stars (He-sdOs). We analysed multiple archival spectra of the three systems and derived their stellar properties. We find that they all lack radial velocity variability, suggesting formation via a merger channel. However, we derive higher than typical hydrogen abundances for their spectral type, which are in disagreement with current model predictions. Our findings suggest a lower limit to the magnetic fraction of hot subdwarfs of $0.147^{+0.143}_{-0.047}$ per cent, and provide evidence for merger-induced magnetic fields which could explain white dwarfs with field strengths of $50-150$~MG, assuming magnetic flux conservation.
\end{abstract}




\section{Introduction}

Magnetic fields have been detected in stars across many evolutionary stages, from the main sequence \citep{Babcock1947} to the white dwarf cooling sequence \citep{Kemp1970}, since many decades. Yet the origin and evolution of these fields is not entirely understood \citep[e.g.][]{Ferrario2015,Wurster2018}. For white dwarfs, the final observable evolutionary stage of over 95 per cent of stars, the fraction of systems with detectable magnetic fields is estimated to be over one fifth \citep[$22\pm4$ per cent,][]{Bagnulo2021}.

Several mechanisms have been put forward to explain the magnetic fields observed in white dwarfs. Firstly, the magnetic field could be explained simply as a fossil field that was already present in the cloud from which the star originally formed \citep{Woltjer1964, Landstreet1967, Angel1981}. In this scenario, the field strength results from flux conservation when the progenitor star contracts to become a white dwarf, with magnetic Ap and Bp stars \citep{Moss2001} being the likely progenitors of magnetic white dwarfs. Alternatively, the fossil field could arise due to a dynamo acting in the convective core during the main sequence or the asymptotic giant branch \citep{Stello2016} and only be revealed after the white dwarf progenitor loses its outer layers. Another model suggests that the magnetic field could result from a dynamo generated during the merger of two stars forming a white dwarf \citep{Tout2008, Briggs2015, Briggs2018}, or from the merger of two white dwarfs \citep{GarciaBerro2012}. A merger during an earlier evolutionary stage \citep[the main sequence or even pre-main sequence,][]{Ferrario2009, Schneider2016, Schneider2019} leading to a magnetic main sequence star that evolves to a magnetic white dwarf is also a possibility. Finally, another scenario proposes that the magnetic fields in white dwarfs are generated during the cooling of the star itself \citep{Valyavin1999}, for example due to crystallisation, which induces the formation of a convective mantle around the solid white dwarf core \citep{Isern2017}. However, none of these scenarios alone can fully explain the observed fraction and field strengths of magnetic white dwarfs; likely more than one scenario is required \citep{Bagnulo2021}.

Before reaching the white dwarf stage, a small fraction of systems will go through the extended horizontal branch (EHB), where they are referred to as hot subdwarf stars \citep[see][for a review]{Heber2016}. These stars appear hot and smaller than canonical horizontal branch stars due to previous enhanced mass-loss attributed to binary interaction \citep{Han2002, Han2003, Pelisoli2020}. They will evolve directly to the white dwarf cooling track without ascending the asymptotic giant branch. Despite this direct connection with white dwarfs, the fraction of magnetic hot subdwarfs seems to be much smaller than that of magnetic white dwarfs. Searches using spectropolarimetry found no evidence of magnetic fields in around 40 hot subdwarfs, even with detection limits as low as 1 to 2~kG \citep{Landstreet2012,Mathys2012,Randall2015,Bagnulo2015}. The picture is not much better for detection through Zeeman splitting: to date, out of around 6000 spectroscopically confirmed hot subdwarfs \citep{Geier2020, Culpan2022}, there is only one confirmed magnetic hot subdwarf \citep{Dorsch2022}. An earlier work by \citet{Heber2013} claimed a first detection and reported a magnetic field strength of 300--700 kG from Zeeman-split hydrogen and helium lines, but the reported star was never named or analysed in detail. In addition, the merger remnant J22564-5910 could host a magnetic field, but the observed spectral features could instead be explained by a disc \citep{Vos2021}. The detection of photometric variability consistent with spots could point towards a magnetic field for a number of hot subdwarfs \citep{Jeffery2013,Geier2015,Balona2019,Momany2020}, but the cause for variability and its possible connection to a magnetic field remains to be investigated. This conflict between an abundance of magnetic white dwarfs and a dearth of magnetic hot subdwarfs might contain clues about the possible channels leading to the formation of magnetic white dwarfs, and thus to the behaviour of magnetic fields throughout stellar evolution, calling for more investigation of possible magnetic fields in hot subdwarfs.

In this work, we report the discovery and characterisation of three magnetic hot subdwarfs: SDSS~J041536.05+253857.1
SDSS~J130346.61+264630.6, and SDSS~J160325.52+341237.4 (henceforth \jofour, \jthirteen, \jsixteen, respectively). This discovery represents a significant increase in the number of known magnetic hot subdwarfs, and can shed light onto the origin and evolution of stellar magnetic fields.

\section{Spectroscopic and photometric data}

\begin{figure*}
	\includegraphics[width=\textwidth]{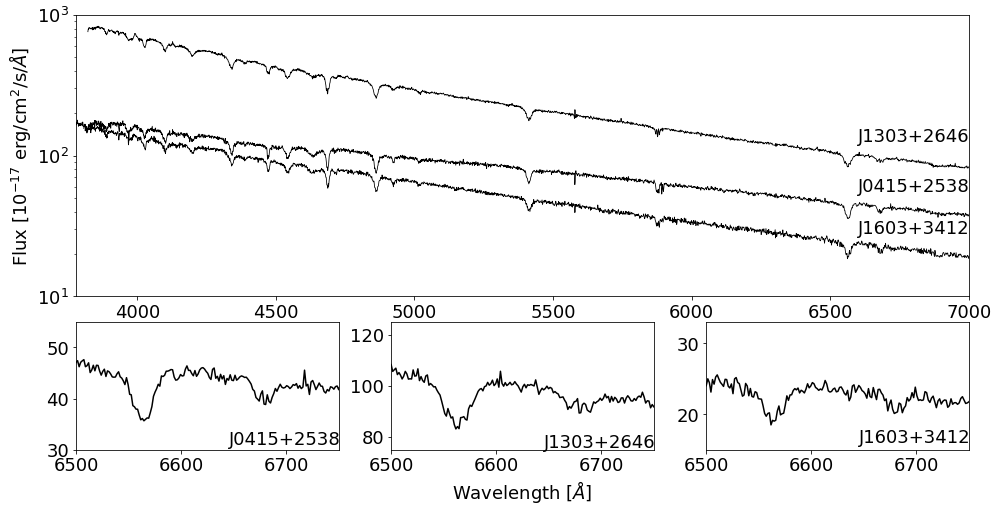}
    \caption{SDSS spectra of \jofour, \jthirteen, and \jsixteen\ are shown in the top panel. The bottom panel zooms in the region around H~$\alpha$ and the He~{\sc i} 6678~\AA\ line, which show hints of Zeeman splitting. \jofour\ is in a region with strong reddening (see Sec.~\ref{sec:specfit}).}
    \label{fig:SDSS}
\end{figure*}

We identified the possible presence of a magnetic field in the three stars based on visual analysis of spectra taken with the Sloan Digital Sky Survey \citep[SDSS,][]{sdssiii}. The three targets were part of a sample of candidate white dwarfs identified by their colours, but were instead found to show narrower lines and very blue spectra consistent with hot subdwarfs (see Fig.~\ref{fig:SDSS}). The strength of the helium lines compared to the hydrogen lines and the presence of He~{\sc ii} lines imply a He-sdO classification for all three objects. In addition, we identified hints of Zeeman splitting of the Balmer lines, caused by the magnetic field breaking azimutal symmetry.

We then searched the database of the Isaac Newton Group of telescopes\footnote{http://casu.ast.cam.ac.uk/casuadc/ingarch/query} for available spectroscopy for the three objects. We found multiple archival spectra taken with the Intermediate-dispersion Spectrograph and Imaging System\footnote{https://www.ing.iac.es/Astronomy/instruments/isis/} (ISIS) at the William Herschel Telescope (WHT). Data from seven nights was available for \jofour\ (Table~\ref{tab:data04}), three nights for \jthirteen\ (Table~\ref{tab:data13}), and two for \jsixteen\ (Table~\ref{tab:data16}). In most cases, more than one spectrum was taken each night. For all observations, except those taken on 2015 December 15 for \jofour, arc lamps were taken in the same position as the target.

\begin{table}
	\centering
	\caption{List of archival WHT/ISIS spectra retrieved for \jofour}
	\label{tab:data04}
	\begin{tabular}{ccccc} 
		\hline
		Date & Grating & \multicolumn{2}{c}{Central wavelength (\AA)} & Number of spectra\\
		     &         & Blue & Red & \\
		\hline
		20140203 & R600 & 4300 & 6403 & 2\\
		20140204 & R600 & 4300 & 6403 & 2\\
        20150822 & R600 & 4298 & 6201 & 4\\
        20150823 & R600 & 4298 & 6201 & 4\\
        20150824 & R600 & 4298 & 6201 & 4\\
        20150825 & R600 & 4298 & 6201 & 4\\
        20151215 & R600 & 4498 & 6900 & 3\\
        \hline
	\end{tabular}
\end{table}

\begin{table}
	\centering
	\caption{List of archival WHT/ISIS spectra retrieved for \jthirteen}
	\label{tab:data13}
	\begin{tabular}{ccccc} 
		\hline
		Date & Grating & \multicolumn{2}{c}{Central wavelength (\AA)} & Number of spectra\\
		     &         & Blue & Red & \\
		\hline
        20050225 & R1200 & 4501 & 6199 & 1\\
        20120531 & R600 & 4351 & 6558 & 4\\
        20150615 & R1200 & 4750 & 6799 & 4\\
        20150616 & R1200 & 4750 & 6799 & 6\\
        \hline
	\end{tabular}
\end{table}

\begin{table}
	\centering
	\caption{List of archival WHT/ISIS spectra retrieved for \jsixteen}
	\label{tab:data16}
	\begin{tabular}{ccccc} 
		\hline
		Date & Grating & \multicolumn{2}{c}{Central wavelength (\AA)} & Number of spectra\\
		     &         & Blue & Red & \\
		\hline
        20150615 & R1200 & 4750 & 6799 & 4\\
        20150616 & R1200 & 4750 & 6799 & 5\\
        \hline
	\end{tabular}
\end{table}

We downloaded all the spectra and associated calibration files and performed data reduction and optimal extraction \citep{Marsh1989} using {\sc pamela}\footnote{https://cygnus.astro.warwick.ac.uk/phsaap/software/pamela/html/INDEX.html}. All spectra were de-biased and flat-fielded using the standard {\sc starlink}\footnote{https://starlink.eao.hawaii.edu/starlink} packages {\tt kappa}, {\tt figaro} and {\tt convert}. Wavelength calibration was carried out using {\sc molly}\footnote{https://cygnus.astro.warwick.ac.uk/phsaap/software/molly/html/INDEX.html}.

In order to search for photometric variability in the three stars, in particular variations that could be attributed to spots, we queried the database of the Transiting Exoplanet Survey Satellite \citep[TESS,][]{tess} using the Mikulski Archive for Space Telescopes (MAST). \jofour\ (TIC~56742534) was observed in sectors 43 and 44 with cadences of 20~seconds and 2~minutes, whereas for \jthirteen\ and \jsixteen\ only 30-minute full-frame images (FFIs) are available during one and two sectors, respectively.

\begin{figure}
	\includegraphics[width=\columnwidth]{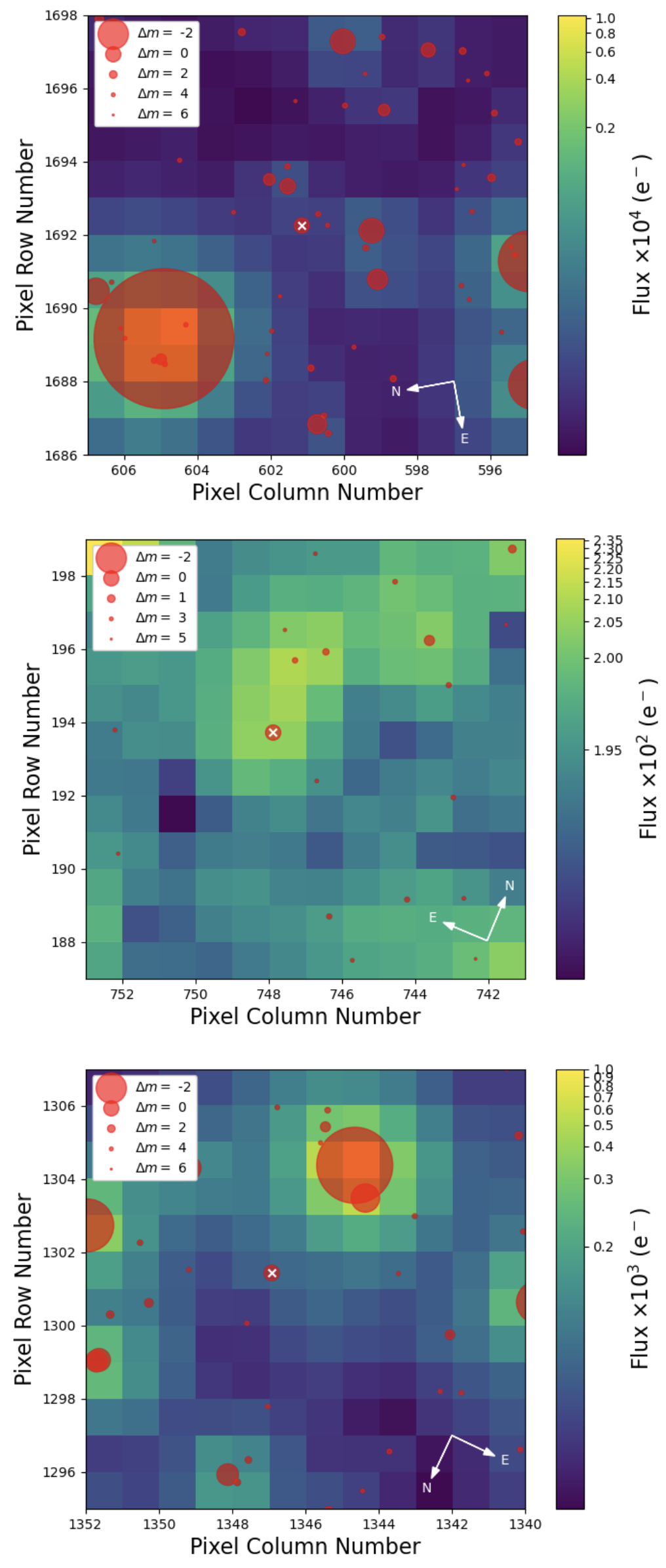}
    \caption{TESS field of view for the three targets, \jofour, \jthirteen, and \jsixteen\ from top to bottom. The targets are marked by a white cross, and other stars in the field with a magnitude difference ($\Delta m$) of up to six are also indicated. Both \jofour\ and \jsixteen\ have bright stars nearby that likely contaminate their TESS light curves. Images generated with {\tt tpfplotter} \citep{tpfplotter}.}
    \label{fig:tpf}
\end{figure}

Though the cadence and duration of the TESS light curve is adequate for detecting rotation periods typical of most hot subdwarfs \citep[$\lesssim 50$~days][]{Charpinet2018, Reed2018}, rotation periods nearing a hundred days have been detected for some hot subdwarfs \citep{Reed2014, Bachulski2016}. In addition, TESS observations can suffer from significant contamination from nearby stars given the large pixel size of 21~arcsec. In fact, the reported contribution of \jofour\ to the TESS aperture is only 26~per cent. Only \jthirteen\ seems to be fairly isolated, since the TESS observations of \jsixteen\ are also possibly contaminated by a nearby bright star (see Fig.~\ref{fig:tpf}). For these reasons, we have also retrieved light curves from the Zwicky Transient Facility \citep[ZTF,][]{ztf} and the Catalina Real Time Transient Survey \citep[CRTS,][]{crts} for our three targets, given the better spatial resolution and often longer time span of these surveys compared to TESS.

\section{Data analysis}

\subsection{Spectral and spectral energy distribution fitting}
\label{sec:specfit}

The spectral analysis for our three targets was performed following the method used by \citet{Dorsch2022} to model the prototype magnetic He-sdO,  Gaia~DR2~5694207034772278400 (henceforth \joeight). Atmospheric structures were computed using the plane-parallel, homogeneous, and hydrostatic code \tlusty\ \citep{hubeny17b, hubeny17c}, including H, He, C, N, O, Ne, Si, P, S, Fe, and Ni\footnote{Like \cite{Dorsch2022}, we used high abundances for iron (1.5 times solar) and nickel (10 times solar), as well as a high microturbulence (5\,\kms) to approximate the additional opacity due to Zeeman splitting in the far-ultraviolet spectral region.} in non-local thermodynamic equilibrium. The magnetic field was not considered in the atmospheric structure and only linear Zeeman splittings were included in the spectrum synthesis, which was performed with \textsc{Synspec} \citep{hubeny17a}. A simple homogeneous and uniform magnetic field across the visible hemisphere was assumed. Polarised radiative transfer in the lines was not considered. A more detailed description of our methods is given in section~3 and appendix~B of \cite{Dorsch2022}.

\begin{figure*}
\centering
\includegraphics[width=0.925\textwidth]{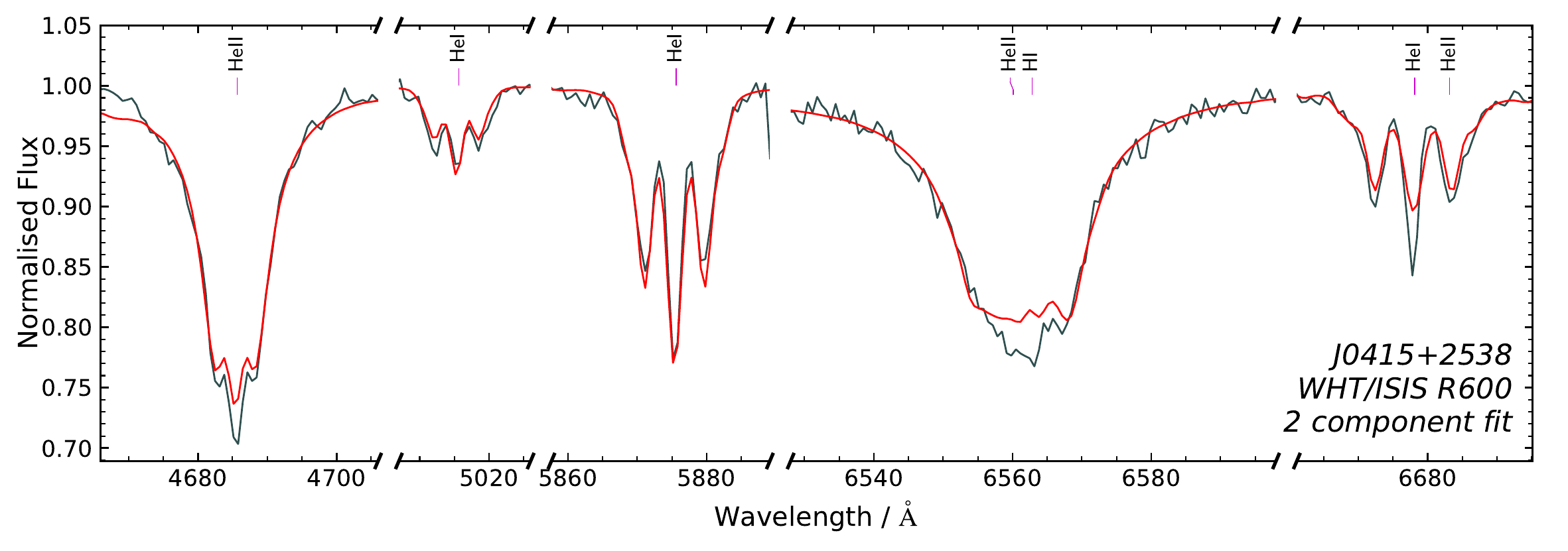}\vspace{-16pt}
\includegraphics[width=0.925\textwidth]{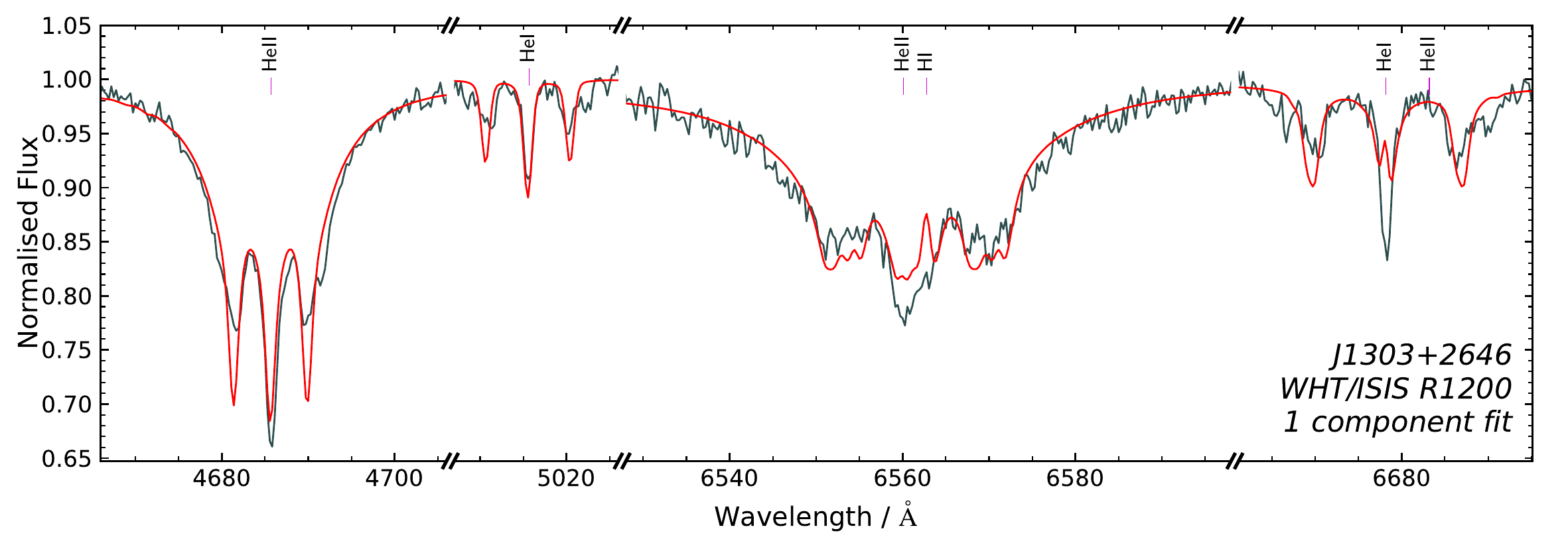}\vspace{-16pt}
\includegraphics[width=0.925\textwidth]{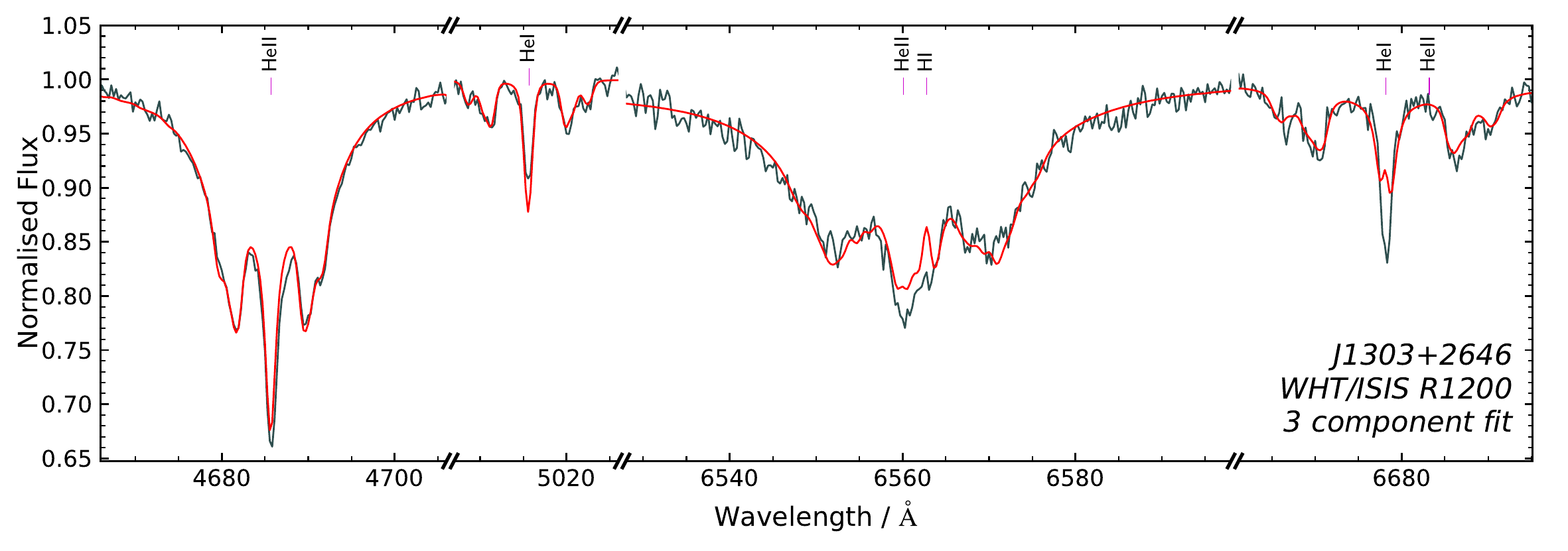}\vspace{-16pt}
\includegraphics[width=0.925\textwidth]{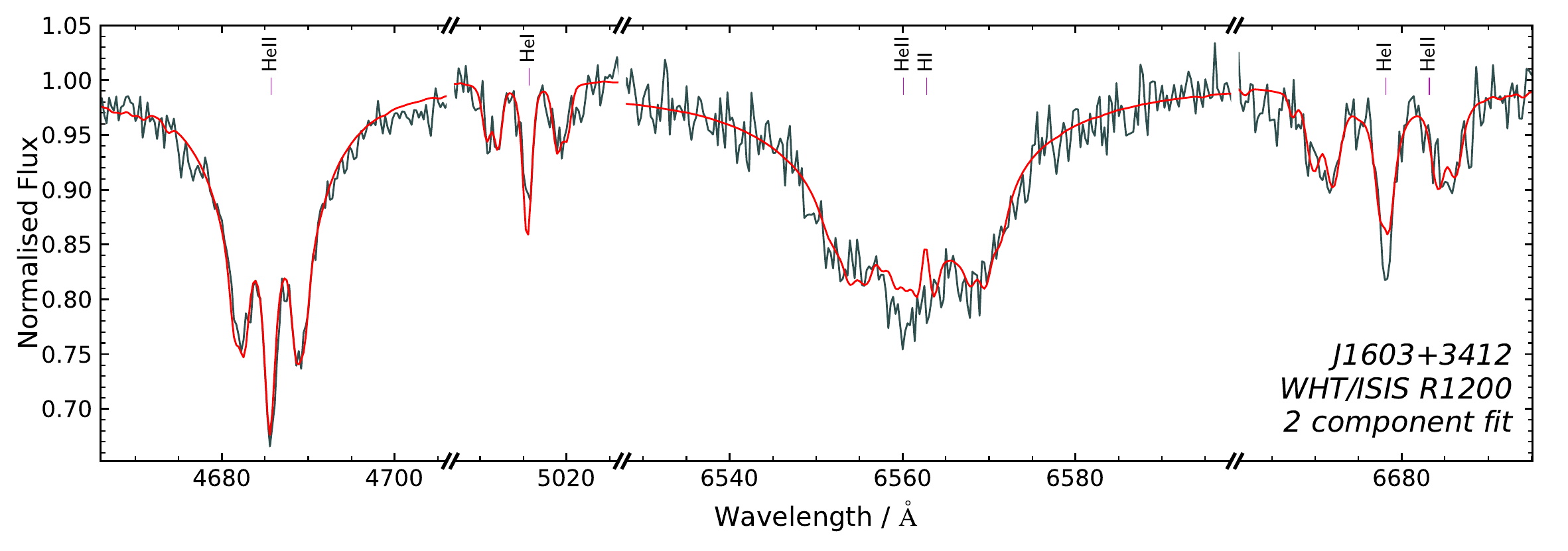}
\caption{
\ion{H}{i}, \ion{He}{i} and \ion{He}{ii} lines in the merged and radial velocity-corrected WHT/ISIS spectra for each target. 
The best model is shown in red, not including metal lines. 
Labels indicate \ion{H}{i} and \ion{He}{i-ii} line positions at $B=0$. The top panel shows our best fit for \jofour. The two middle panels show fits for \jthirteen: initially using only one magnetic field component, which leads to a poor fit to the Zeeman components, and using three components, which can much better approximate the complex magnetic field geometry. The bottom panel shows the final fit for the merged spectrum of \jsixteen.
}
\label{fig:atmlines}
\end{figure*}

We performed global $\chi ^2$ fits to the WHT/ISIS spectra of each star. Initially we fitted the Doppler-corrected co-added spectra to evaluate the performance of our simple treatment of the magnetic field. The free parameters were the effective temperature \teff, the surface gravity \logg, the helium abundance \logy, and the mean magnetic field strength $B$. This initial fit showed that the spectra of \jthirteen\ clearly display broadened displaced Zeeman components (see Fig.~\ref{fig:atmlines}), which indicates that the magnetic field across the surface of this star is non-homogeneous. To account for that, we constructed toy models consisting of more than one homogeneous component, which allowed us to roughly emulate a non-homogeneous magnetic field geometry causing variation of the magnetic field strength on the stellar surface. For each star, we re-fitted the co-added spectra with one and two additional homogeneous magnetic field components that were allowed to vary in strength and surface ratio. The results of this exercise are summarised in Table~\ref{tab:components}. 
Importantly, our toy model also allowed us to investigate the systematic uncertainties of the derived atmospheric parameters caused by our approximation of an uniform magnetic field.
The resulting \teff\ values change insignificantly, because they are dominantly constrained by the helium ionisation equilibrium rather than by the detailed spectral line shapes. The surface gravities as well as the hydrogen to helium ratios, however, are derived mainly from the shapes of the hydrogen and helium lines. Therefore, changes of 0.1--0.2~dex are observed when introducing a second component. Adding a third one leads to considerably smaller changes of the atmospheric parameters, which we judge to be insignificant for \jofour\ and \jsixteen, for which we therefore adopted the two-component model. The field structure of \jthirteen\ is more complex, which led us to adopt three components.

Once the number of components was fixed, all available spectra were fitted simultaneously with the selected number of components to determine \teff, \logg, \logy, mean magnetic field strength $B$ and surface ratio $A$ of each component, and the radial velocities \vrad. We only allowed \vrad\ to be different for the individual spectra, forcing a global best-fit for the atmospheric parameters. The magnetic field axis was forced to be inclined at an angle $\psi = 90^\circ$ with respect to the line of sight because our simplified model for the magnetic field geometry does not allow for a physical interpretation of this angle.  The projected rotational velocity was fixed to \vsini\ = 0\,\kms\ for all stars because it is not well constrained by the low-resolution WHT/ISIS spectra. We only derived upper limits based on the value preferred by the fit. Spectral regions that were poorly reproduced by our models were excluded from the fit. This includes \ion{He}{i}\,4471\,\AA, as well as regions that are affected by metal lines.  Important metal line blends are due to strong \ion{N}{iii} lines partly blended with \ion{H}{i}/\ion{He}{ii} 4101, 4862\,\AA\ and \ion{He}{ii} 4201, 4543\,\AA.

\begin{figure}
	\includegraphics[width=\columnwidth]{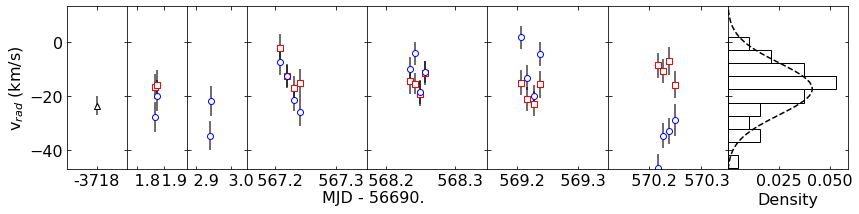}
	\includegraphics[width=\columnwidth]{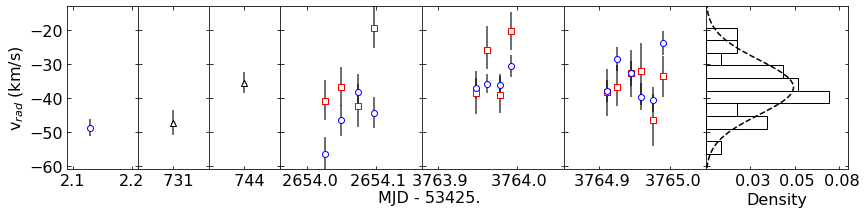}
	\includegraphics[width=\columnwidth]{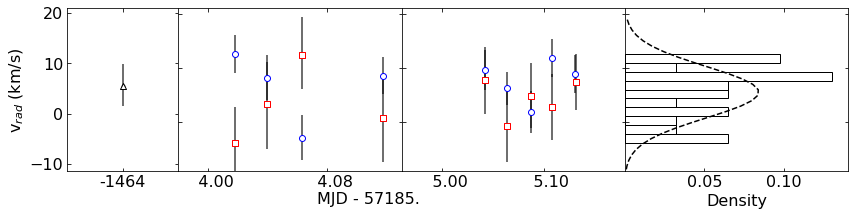}
    \caption{Radial velocities for \jofour, \jthirteen, and \jsixteen, from top to bottom. Estimates obtained from the red and blue WHT arms are shown as red squares and blue circles. Estimates from the SDSS spectra (two available in the case of \jthirteen) are shown as black triangles. The rightmost panel shows a histogram of the values, with a normal distribution with mean and standard deviation derived from the measurements for comparison.}
    \label{fig:vrads}
\end{figure}

Our best-fit models are compared with the merged and radial velocity-corrected WHT/ISIS spectra in Fig.~\ref{fig:atmlines}. The best-fit parameters are listed in Table \ref{tab:results}, which lists the average magnetic field for each star. The strengths and relative surface ratios of the components are given in Table~\ref{tab:components2}. The uncertainties of the atmospheric parameters stated in Table \ref{tab:results} are estimated systematical uncertainties because the statistical uncertainties are negligible in comparison. For the radial velocities, we state the average values and their standard deviations. For \jofour, we exclude the radial velocity measurements taken on 2015 December 15, given that no arc lamp was taken with the same pointing as the target, making the radial velocities unreliable due to instrumental shifts. In all three cases, there is no evidence of significant radial velocity variability in timescales spanning thousands of days (see Fig.~\ref{fig:vrads}), comparable to the longest orbital periods observed for hot subdwarfs \citep{Vos2019}, indicating that the three stars are single.

\begin{table*}
\caption{Stellar parameters derived from spectroscopic and SED fits. We include also the values for the prototype star \joeight\ from \citet{Dorsch2022} for comparison. For \teff, \logg, and \logy, we quote the systematic uncertainties which are dominant over the statistical ones. For \vrad, we quote the average and standard deviation over the multiple measurements. For $R$ and $L$, the quoted values are the mode and the 68 per cent confidence interval.}
\label{tab:results}
\begin{center}
\begin{tabular}{l r r r r}
\hline
{} & \joeight\ & \jofour & \jthirteen & \jsixteen   \\
\hline
$T_\mathrm{eff}$ (K) & $44900 \pm 1000$ & $46580 \pm 1500$  & $47950 \pm 1500$ & $46450 \pm 1500$ \\
$\log g$ & $\phantom{+}5.93 \pm 0.15$ & $\phantom{+}5.98 \pm 0.25$   & $\phantom{+}5.97 \pm 0.30$ &  $\phantom{+}6.06 \pm 0.20$ \\
$\log n(\mathrm{He}) / n(\mathrm{H})$ & $+0.28 \pm 0.10$ & $-0.10 \pm 0.15$   & $+0.25 \pm 0.15$ & $+0.07 \pm 0.15$ \\
$B_\mathrm{avg}$ (kG) & $353 \pm 10$ & $305 \pm 20$ & $450 \pm 20$ & $335 \pm 15$ \\
$v_\mathrm{rad}$ (km\,s$^{-1}$) & $33\pm2$  & $-17\pm10$ &  $-37\pm8$ & $6\pm5$ \\
$v_\mathrm{rot}\sin i~$ (km\,s$^{-1}$) & $<40$ & $<45$ & $<60$ & $<65$ \\[2pt]
$R$ (R$_{\sun}$) & $0.184^{+0.011}_{-0.010}$  &  $0.148^{+0.020}_{-0.015}$  & $0.19^{+0.05}_{-0.04}$   &  $0.14^{+0.06}_{-0.04}$  \\[2pt]
$L$ (L$_{\sun}$) & $123^{+19}_{-16}$  &  $91^{+29}_{-21}$  & $160^{+100}_{-60}$   &  $70^{+80}_{-40}$ \\
\hline
\end{tabular}
\end{center}
\end{table*}

\begin{figure}
	\includegraphics[width=0.99\columnwidth]{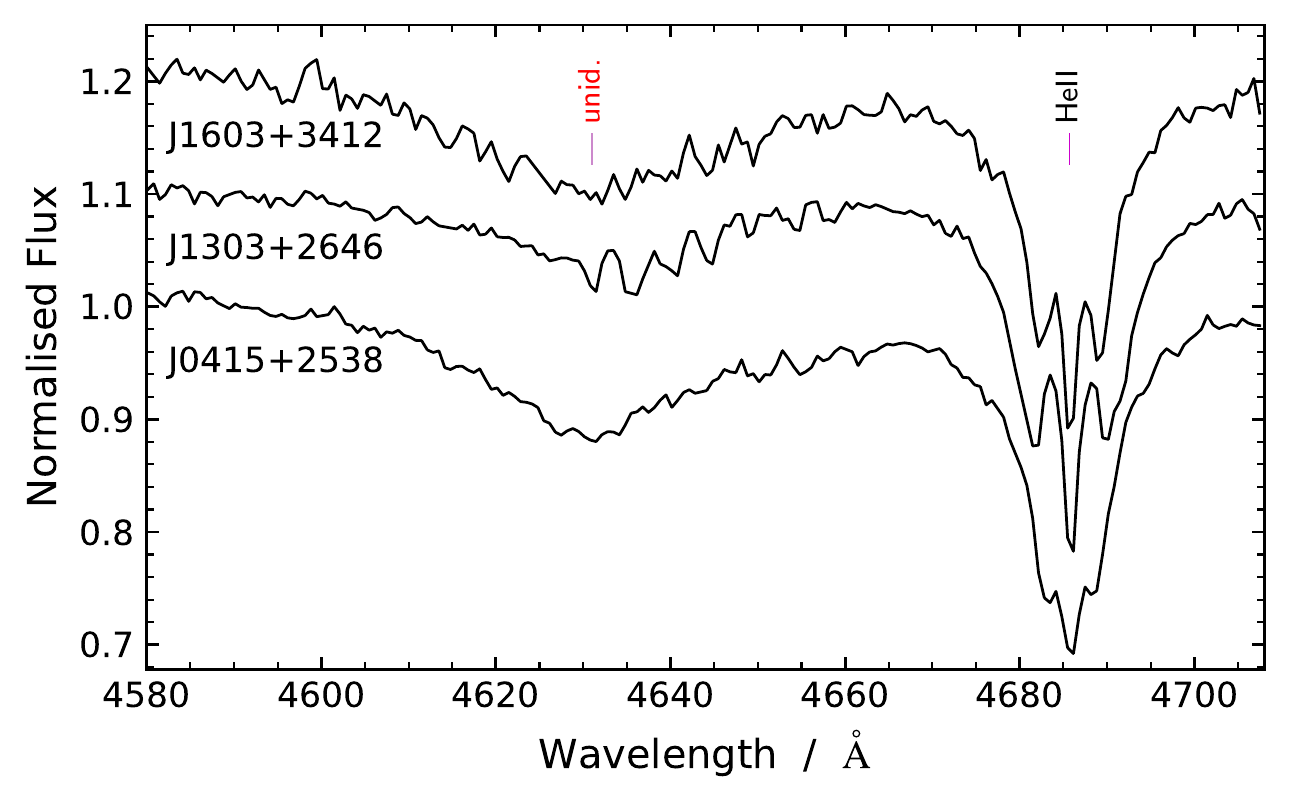}
    \caption{Merged and radial velocity-corrected WHT/ISIS spectra from top to bottom for \jsixteen, \jthirteen, and \jofour. The spectra are offset in steps of 0.1 for better visibility. The origin of the broad and smooth feature centred at about 4631\,\AA\ is unknown. 
    }
    \label{fig:feature}
\end{figure}

The similarities between the atmospheric parameters of all four known magnetic He-sdOs are remarkable. 
All stars share an intermediate helium abundance, with almost the same number of hydrogen and helium atoms in their photospheres. This is highly unusual for He-sdO stars at \teff\ > 43\,000\,K, which are almost always extremely hydrogen-poor or helium-poor \citep{Stroeer2007,Luo2021}. The distinction of two groups of He-sdOs based on hydrogen abundance was suggested by \citet{Naslim13}, who named those with significant hydrogen (H/He $>$ 0.25), like our objects, intermediate He-sdO (iHe-sdO). Those with lower hydrogen content are called extreme He-sdO (eHe-sdO). An additional subdivision was proposed by \citet{Stroeer2007} and \citet{hirsch09}, who demonstrated that the He-sdOs from the ESO supernovae type Ia progenitor survey (SPY) project can be split into four groups characterised by their carbon and nitrogen content: N-rich, C-rich, C\&N-rich, and N-poor objects. Due to the low resolution of the available spectra, detailed abundance patterns could not be determined. 
All stars seem to lack strong carbon lines, similar to \joeight.
Hints of the \ion{C}{iv} lines at 5805\,\AA\ and the \ion{C}{iii}\,4070\,\AA\ triplet are observed in the merged WHT/ISIS spectrum of \jofour\ and to a lesser degree in the SDSS spectrum of \jsixteen, but are absent in the WHT/ISIS spectrum of \jthirteen.
This suggests that carbon is not strongly enriched, although solar carbon abundances cannot be excluded. 
The \ion{N}{iii}\,4517, 4639\,\AA\ multiplets in the WHT/ISIS spectra of \jthirteen\ are best reproduced at a nitrogen abundance of about ten times solar.
The same lines are weaker in the spectra of \jofour\ and \jsixteen, suggesting nitrogen abundances between two and six times solar. In short, there is indication that the magnetic objects are N-rich, but better spectra are needed to probe the C content.

In addition, all stars show a strong and broad feature in the 4629 -- 4660\,\AA\ range, centred at about 4631\,\AA\ (see Fig.~\ref{fig:feature}). The origin of the feature remains unclear.
A photospheric origin seems to be excluded by the lack of similar features at other wavelengths. 
The same argument can be used to exclude both ultra-high excitation lines, which are observed for some DO-type white dwarfs \citep{Werner1995,Reindl2019}, and diffuse interstellar bands. 
An instrumental effect is excluded because the feature is also observed in the SDSS spectra.
The feature is present in the X-SHOOTER spectrum \joeight\ as well, but weaker than in the three new stars.

Following \citet{Dorsch2022}, we also fitted the spectral energy distribution (SED) of the three stars using the same model grid.  The SED was constructed by collecting photometric measurements from multiple surveys (see Appendix~\ref{sec:phot}). \teff, \logg, and \logy\ were fixed to the values determined from spectroscopy, and the angular diameter $\Theta$ was left as a free parameter. We used the law of \citet{Fitzpatrick2019} to account for interstellar extinction, with the colour excess $E_{44-55}$ left to vary freely, but keeping a fixed extinction parameter $R(55) = 3.02$. We combined the derived $\Theta$ with the parallax from {\it Gaia} EDR3 \citep{Gaia2016, Gaia2020} to estimate the stellar radii $R$ and luminosities $L$. We applied a parallax correction to the parallax following \citet{Lindegren2021}, and inflated its uncertainty according to equation 16 of \citet{El-Badry2021}. In principle, the stellar mass could be determined from the radius and \logg\ measurements, but the large uncertainties preclude any meaningful results. The obtained radii and luminosities are listed in Table \ref{tab:results}. Although these luminosities are higher than for canonical sdB hot subdwarfs, they are consistent with what has been previously derived for He-sdOs \citep[see e.g.][]{Stroeer2007}. We find a significant reddening of $E_{44-55} = 0.298\pm0.005$~mag for \jofour, in agreement with reddening maps \citep[e.g.][]{Lallement2018}, whereas \jthirteen\ and \jsixteen\ are not strongly reddened ($E_{44-55} = 0.0049\pm0.0028$~mag and $E_{44-55} = 0.025\pm0.006$~mag, respectively).

\subsection{Light curve analysis}

\begin{figure*}
	\includegraphics[width=\textwidth]{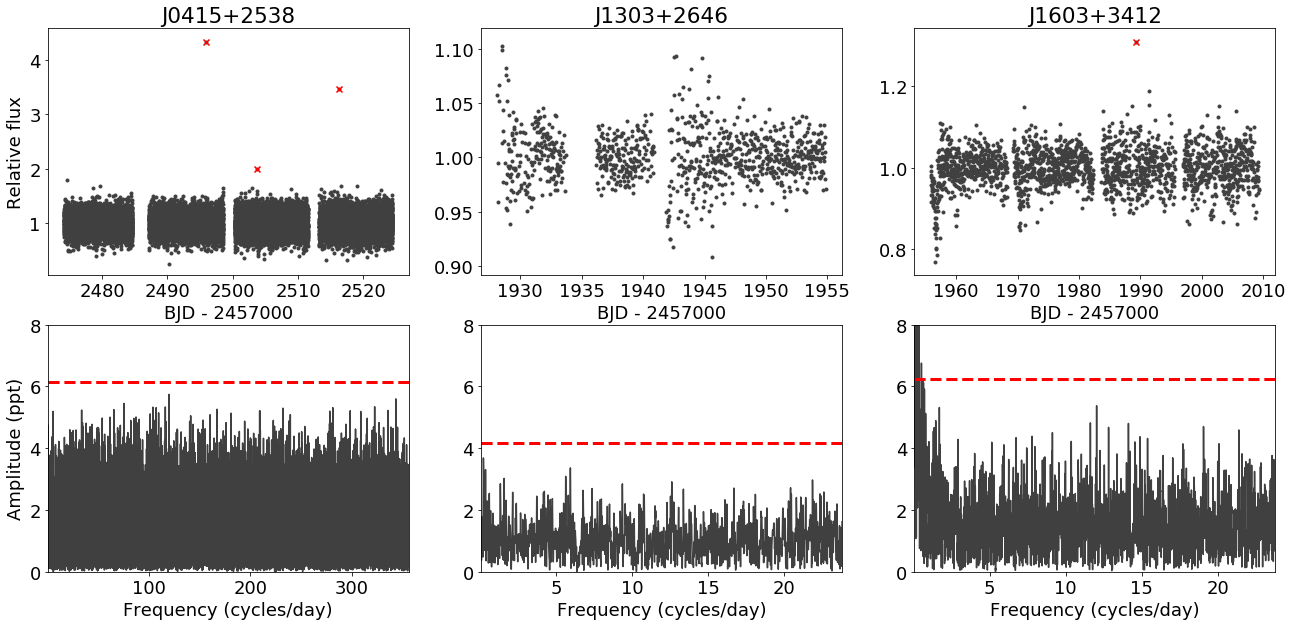}
    \caption{The top panels show the TESS light curves for our three targets as indicated. Points excluded from the analysis are marked by crosses. The bottom panels show the respective Fourier transforms, with the dashed line indicating an adopted detection limit of four times the average amplitude. Aside from low-frequency noise in the periodogram of \jsixteen, no significant peaks appear.}
    \label{fig:tess}
\end{figure*}

We retrieved the light curves for \jofour\ provided by the TESS Science Processing Operations Center (SPOC) pipeline. Given the range of periods in which we are interested, we focus the analysis on the 2-minute light curve, which provides a better signal-to-noise ratio. For \jthirteen\ and \jsixteen, for which no SPOC light curves are available, we used {\sc eleanor} \citep{Feinstein2019} to perform the photometry. We excluded from the analysis any points more than five standard deviations away from the median, and calculated a Fourier transform for each light curve up to the Nyquist frequency. Light curves and periodograms are shown in Fig.~\ref{fig:tess}.

For ZTF and CRTS, we retrieved the light curves from their respective databases for each of our targets. In the case of ZTF, there are two different filters available, $r$ and $g$, and both were retrieved. A Fourier transform was calculated in the same way as for the TESS data, with the Nyquist frequency estimated from the median cadence of observations taken on the same night. Results for ZTF and CRTS are shown in Appendix~\ref{sec:lcs} (Figs.~\ref{fig:ztf} and \ref{fig:crts}, respectively).

We do not identify any signs of periodic variability for our targets. The few possibly significant peaks that appear in the Fourier transforms are either multiples of one-day aliases, given the nightly observations of ZTF and CRTS, or appear marginally above the threshold only for one survey and not the others. We can rule out periodic variability in the range of a few minutes to $\approx 600$~days down to an amplitude of 0.6 per cent for \jofour\ based on the TESS and ZTF light curves, and even longer periods of up to $\approx 1000$~days are ruled out by CRTS down to $\approx 1.5$~per cent. For \jthirteen, TESS rules out periods between an hour and 13~days with amplitudes larger than $\approx 0.4$~per cent, whereas CRTS rules out periods up to $\approx 1000$~days down to $\approx 1.2$~per cent (the ZTF light curve is in turn quite scarce for this object). Finally, for \jsixteen, TESS and ZTF rule out periods between an hour and $\approx 600$~days down to $\approx 0.5$~per cent, whereas the CRTS light curve is not particularly constraining given that the magnitude of the target is near the CRTS detection limit.

\section{Discussion}

\subsection{The detection of magnetic fields in hot subdwarfs}\label{sec:magnetic}

Our three new detections increase the number of hot subdwarfs with confirmed magnetic fields from one to four\footnote{The object mentioned by \citet{Heber2013} is in fact part of our sample.}. Considering that there are 2036 hot subdwarfs identified from SDSS spectra \citep{Geier2020}, and assuming that there is no bias in selecting magnetic systems (which is reasonable since their colours do not seem to be strongly affected), the three detections from SDSS spectra imply a lower limit to the magnetic fraction of hot subdwarfs of $0.147^{+0.143}_{-0.047}$ per cent. Given the low-resolution of SDSS ($R \approx 2000$), only field strengths larger than $\sim 200$~kG can be identified from visual inspection, implying that lower fields would remain undetected. This detection limit is significantly improved for high resolution ($R\approx20000$), which would reveal fields down to $\sim 50$~kG. However, high resolution spectra are available for a smaller number of stars ($\approx 200$) which are not homogeneously selected.

Previous searches for magnetic fields in hot subdwarfs mainly used low-resolution spectropolarimetry \citep{Landstreet2012,Mathys2012}, which has the advantage of lower detection limits of the order of a few hundred gauss to kilogauss, but the disadvantage of requiring the targets to be fairly bright. These searches targeted forty stars of quite different spectral types in various stages of stellar evolution, including sdB stars in close binary systems with white dwarfs as well as low-mass main sequence companions (see Appendix~\ref{sec:specs}). Most observations were carried out with the FORS spectropolarimeter at the ESO VLT. \citet{Landstreet2012} and \citet{Bagnulo2012} reanalysed most FORS observations of hot subdwarfs and found no detections even at 2$\sigma$ level, concluding that there is ``no evidence for the presence of magnetic fields at the level of 1 kG''.

There are five He-sdOs that have been probed by spectropolarimetry, two eHe-sdO stars and three iHe-sdO stars. \citet{Landstreet2012} derived a mean $B_z = 90\pm140$~G for the eHe-sdO CD-31~4800 and $B_z = 232\pm178$~G for the iHe-sdO HD~127493. \citet{Randall2015} reported an upper 3 $\sigma$ limit of 300~G for a magnetic field of the iHe star LS IV$-$14~116. Hence, no magnetic fields at a level of a few hundred gauss are present in these three He-sdOs. Earlier work by \citet{Elkin1996} targeted the eHe-sdO star BD$+$25 4655 and the iHe-sdO BD$+$75 325. They measured circularly polarised spectra using the 6-metre telescope at the Russian Academy of Sciences Special Astronomical Observatory and determined a magnetic field strength of $B_z = 1680\pm60$~G in BD+75~325. Three additional measurements of BD+75~325 pointed at a variable field strength \citep{Elkin1998}. In addition, \citet{Elkin1998} failed to detect a magnetic field at the 400~G level from three observations of BD+25~4655. Hence, BD+75~325 would be the only hot subdwarf with a detected magnetic field of a few kG. However, \citet{Landstreet2012} argue that the real uncertainties in these measurements are likely of the order of 1~kG, i.e. of the same order of the reported fields, hence confirmation would be needed with more sensitive methods. In summary, the fields of the four confirmed magnetic He-sdOs are larger by a factor of at least a thousand than those of the few probed He-sdOs.

\begin{figure}
	\includegraphics[width=\columnwidth]{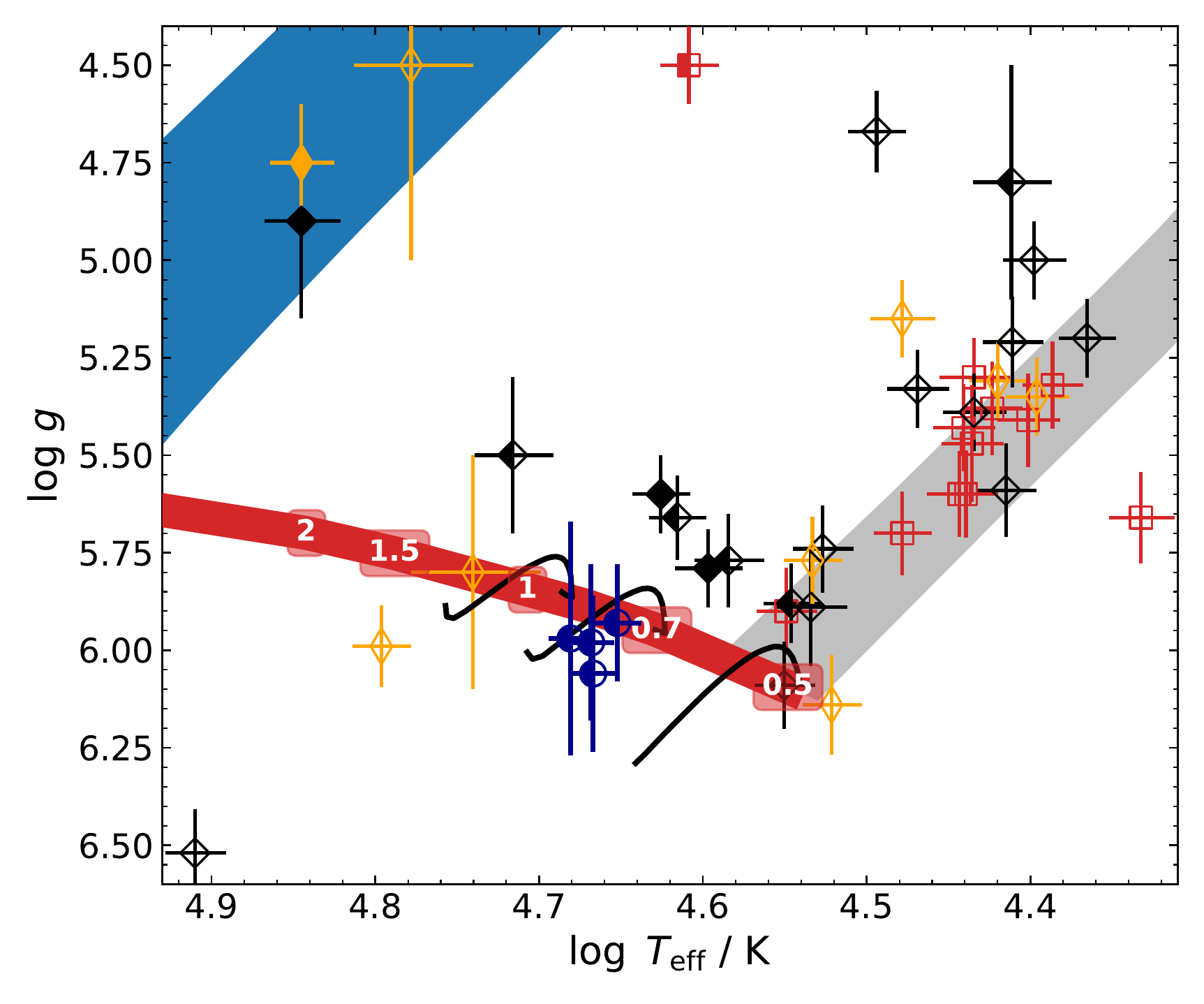}
    \caption{
    Kiel diagram showing hot subdwarf stars in which magnetic fields have been probed for. The four known magnetic He-sdOs are shown as blue circles. Black diamonds mark apparently single (non-\vrad\ variable) stars, red squares show known close binaries with white dwarf or low-mass main sequence/brown dwarf companions (\vrad\ variable), and orange thin diamonds indicate unknown \vrad\ variability. 
    Helium-poor stars are marked by open symbols, extremely He-rich stars by filled symbols, and intermediately He-rich stars by half filled, half open symbols. For details on the objects, see Appendix \ref{sec:specs} and Table \ref{tab:litsds}. The solid black lines indicate the core helium burning phase in the merger tracks of \citet{Yu2021} for a metallicity of $Z=0.01$ and remnant masses of 0.45, 0.65, 0.85\,\msun.
    The grey shaded region marks the location of the EHB by \citet{Dorman1993} for solar metallicity, the blue shaded region marks the range of post-asymptotic giant branch (AGB) tracks of \citet{MillerBertolami2016}, and thick red line indicates the zero age helium main sequence from \citet{Paczynski1971}.
    }
    \label{fig:kiel_magnetic}
\end{figure}

We compare the location of all subdwarfs probed for magnetic fields in the Kiel diagram with the four magnetic He-sdOs in Fig.~\ref{fig:kiel_magnetic}. The binary status of the stars, inferred from \vrad\ variability, is also indicated, as well as the He-enrichment. About 60 per cent of the previously studied stars with sufficient \vrad\ measurements show no evidence of a binary companion, like the known magnetic systems. Strikingly, the four stars for which magnetic fields have been detected cluster very closely together in the Kiel diagram, and none of the previously probed stars are found in this region. This might suggest that a very specific formation scenario is required to generate a magnetic field. However, spectropolarimetric searches in a larger number of stars would be required to confirm that magnetism does not occur for hot subdwarfs in other regions of the Kiel diagram.

\subsection{Formation scenarios for magnetic hot subdwarfs}

Interestingly, all four known magnetic systems are of He-sdO spectral type and show remarkably similar atmospheric parameters (see Table~\ref{tab:results}). This strongly suggests that all four stars were formed by the same evolutionary channel. \citet{Dorsch2022} argued that \joeight\ is likely the result of a merger, given the derived atmospheric parameters and metal abundances. The lack of radial velocity variability for the three stars presented here provides further evidence for a merger origin for magnetic He-sdOs, taking into account that hot subdwarfs are not expected to form without binary interaction \citep{Pelisoli2020}. Indeed, evidence is increasing that the majority of He-rich sdO stars result from mergers. While the fraction of hydrogen-rich subdwarfs in close binaries is high \citep[about 50 per cent,][]{maxted01,napi04}, \citet{Geier2022} showed that radial velocity variables are very rare amongst He-sdOs, concluding that they are likely formed by mergers.

Other He-rich hot subdwarfs likely formed by mergers were observed by the SPY survey \citep{napi03,Lisker2005,Stroeer2007,hirsch09}, which obtained high resolution spectra ($R \approx 20000$) of tens of hot subwarfs. More recent spectral analyses of He-rich sdO stars from high resolution spectroscopy have been reported by \citet{Schindewolf2018}, \citet{Naslim13,Naslim20}, \citet{Dorsch2019}, and \citet{Jeffery2021} while \citet{Latour2018} analysed four He-poor sdOs. In addition, for well over a hundred sdB stars, spectroscopic analyses based on even higher resolution spectroscopy are available \citep[e.g.][]{Edelmann2005,Geier2013,Schneider2018}, but no hint for Zeeman broadening has been found in any of them. Finally, \citet{Werner2022} recently found a CO-rich subtype of He-sdOs whose origin has been attributed to mergers \citep{MillerBertolami2022} which also display no Zeeman splitting. This implies that the magnetic fields in the other analysed stars, if existent, must be much weaker than observed for the four magnetic He-sdOs.

\begin{figure}
	\includegraphics[width=\columnwidth]{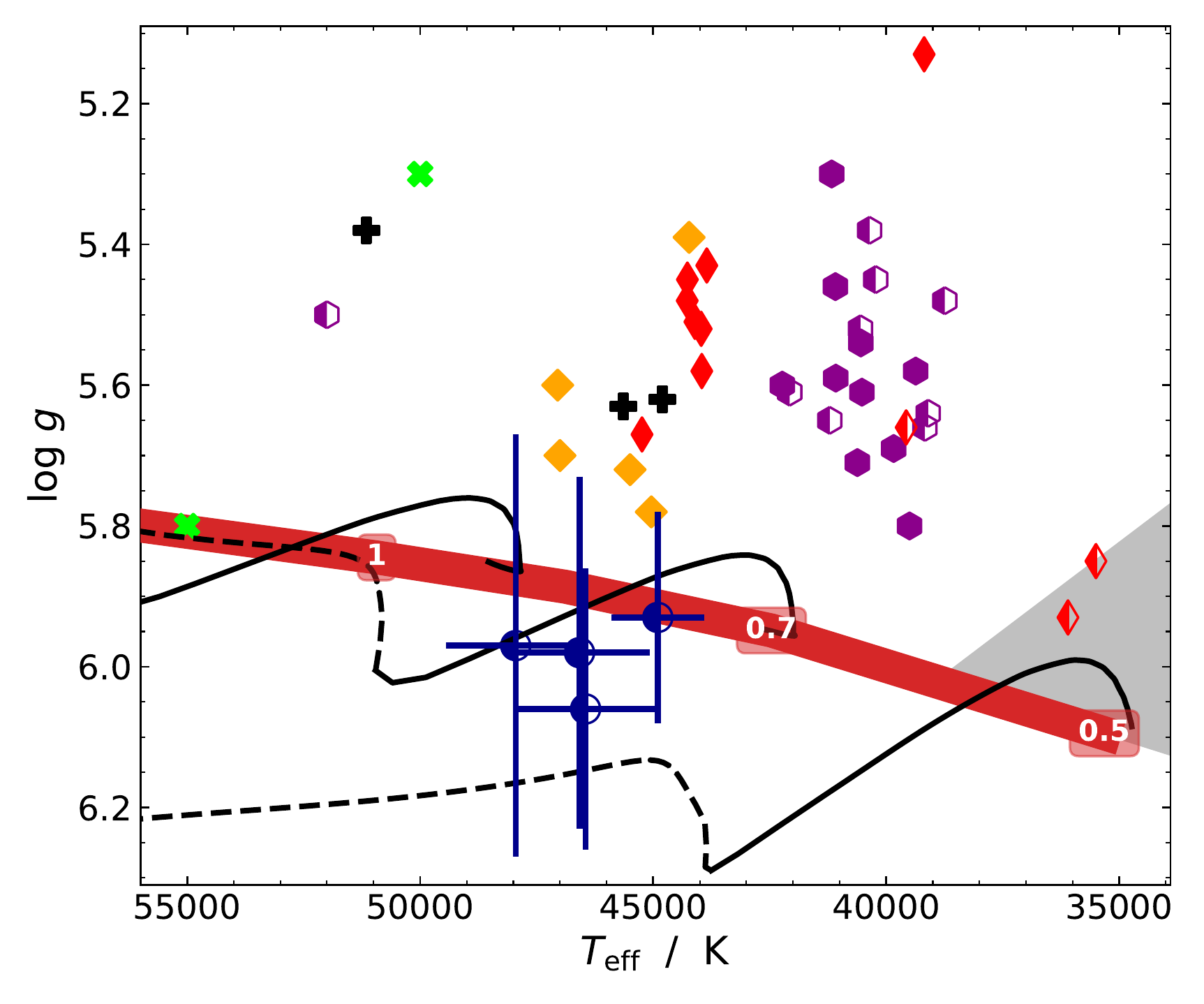}
    \caption{
    Distribution of He-rich hot subdwarf stars in the Kiel diagram. The blue circles with error bars are the magnetic He-sdOs. 
    Extremely He-rich stars are marked by filled symbols and intermediately He-rich stars by half filled, half open symbols. 
    Surface metal abundances are indicated by purple hexagons (N-rich), red thin diamonds (C\&N-rich), orange diamonds (C-rich), or black pluses (C-rich, N-poor). The CO-rich He-sdOs from \citet{Werner2022} are green cross-marks. %
    Merger tracks from \citet{Yu2021} for a metallicity of $Z=0.01$ and remnant masses of 0.45, 0.65, 0.85\,\msun\ are shown as black lines, where solid lines correspond to the core helium burning phase and dashed lines indicate helium shell burning.
    The zero age helium main sequence from \citet{Paczynski1971} is shown as a thick red line.
    The grey shaded region marks the approximate location of the EHB.
    }
    \label{fig:kiel_spy}
\end{figure}

We compare the four magnetic subdwarfs to the He-rich subdwarfs from the SPY project and other detailed high-resolution studies \citep{Lanz1997, Schindewolf2018, Dorsch2019, Dorsch2020, Dorsch2022b},  as well as the CO He-sdOs of \citet{Werner2022} in the Kiel diagram (Fig. \ref{fig:kiel_spy}). The three main subtypes (N-rich, C-rich, C\&N-rich) form two distinct clusters, with the N-rich stars being cooler than the C and C\&N-rich. The two CO-He-sdOs, the three N-poor eHe-sdOs and the four magnetic iHe-sdOs are amongst the hottest He-sdOs.
Though it can be noted that the four magnetic He-sdOs are fairly isolated, it is puzzling that no He-sdO stars other than the four ones discussed here have been found to be magnetic, if mergers were to always lead to magnetic fields. This suggests that some fine-tuning is required in the formation of magnetic systems.

Proposed merger scenarios that could form magnetic hot subdwarfs are the merger of two He-core white dwarfs \citep{Han2003, Zhang2012, Yu2021}, the merger between a hybrid CO/He-core white dwarf and a He-core white dwarf \citep{Justham2011}, and the merger between a He-core white dwarf and a low mass CO-core white dwarf \citep{MillerBertolami2022}. One of the differences between these channels is the resulting mass: the models of \citet{Han2003} and \citep{MillerBertolami2022} can only account for masses up to $\approx 0.8$~M$_{\sun}$, whereas larger masses could be explained by the hybrid merger channel, though the predicted luminosities are higher than those observed for the magnetic He-sdOs. Unfortunately we cannot constrain masses for the studied objects, but future higher-resolution observations and improved astrometry could allow mass estimates to help differentiate between the possible scenarios.

\begin{figure}
	\includegraphics[width=\columnwidth]{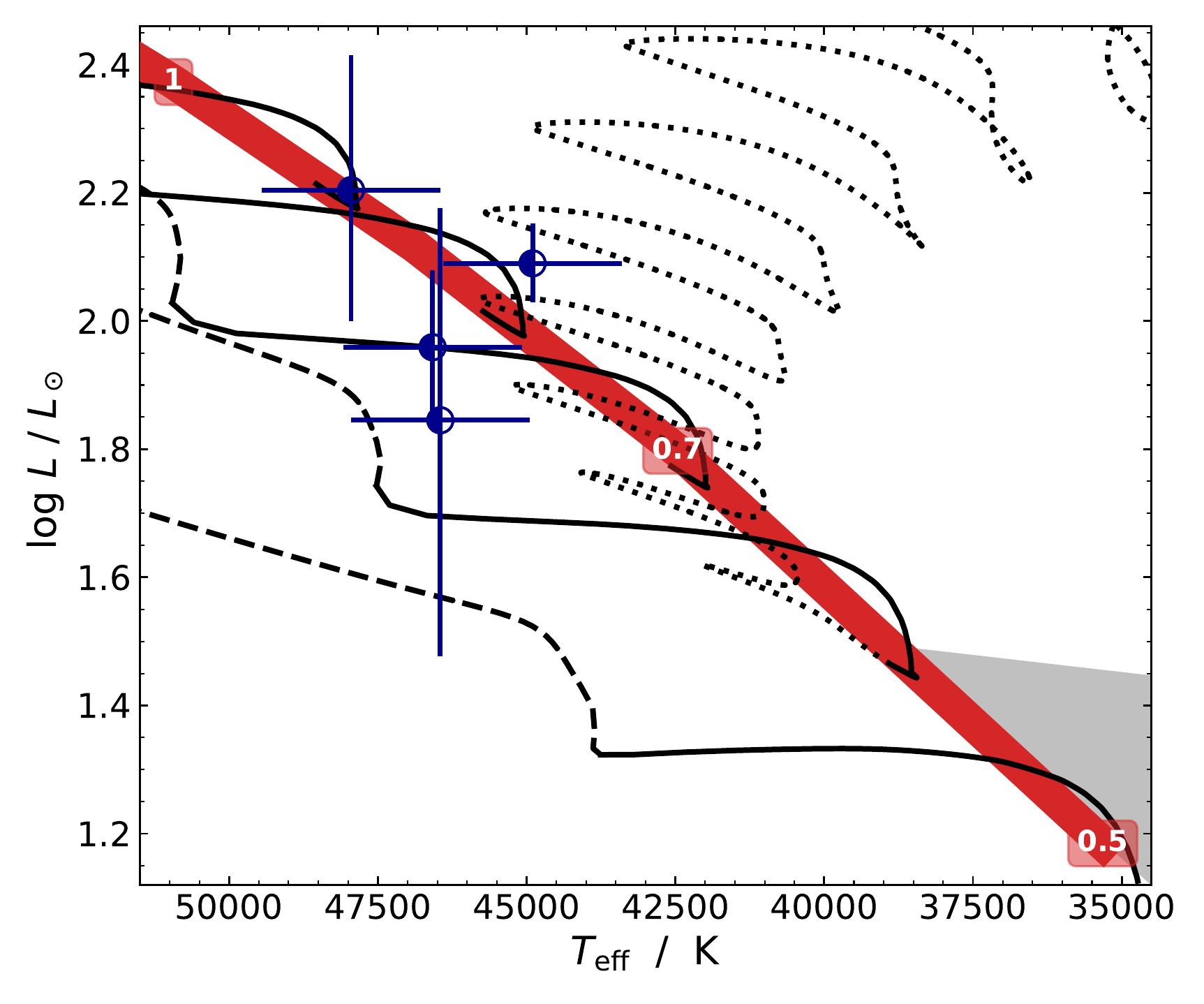}
    \caption{Luminosity as a function of \teff\ for the four magnetic He-sdOs (blue half open dots). 
    Merger tracks from \citet{Yu2021} for a metallicity of $Z=0.01$ and remnant masses of 0.85, 0.75, 0.65, 0.55, and 0.45\,\msun\ are shown in black, where the solid line corresponds to the core helium burning phase and the dashed line indicates helium shell burning. 
    For the 0.55\,\msun\ track, the pre-helium main sequence phase is shown as a dotted line. 
    The broad red line shows the helium zero-age main sequence from \citet{Paczynski1971}, with labelled masses. 
    The grey shaded region marks the approximate location of the EHB. 
    }
    \label{fig:hrd}
\end{figure}

The observed atmospheric abundances can also provide important constraints for the merger models. The rapid mass transfer in He-core white dwarf mergers is predicted to lead to two components \citep{Zhang2012}: a fast accretion event producing a corona around the primary, which is hot enough for helium burning to occur and to produce carbon and convert nitrogen to neon, and a disc from which the material is slowly accreted onto the surface of the primary. The disc is not hot enough to ignite helium burning. Therefore, the composition of the accreted matter is that of the former He-core white dwarf companion, which is He- and N-rich, but C-poor. Composite merger models assume that both components are created in different relative mass fractions. Accordingly, evolutionary calculations of \citet{Zhang2012} predict that C-rich, N-poor surfaces result from fast hot mergers, N-rich surfaces from slow cold mergers and C\&N-rich surfaces from composite models. These variants of the He-core white dwarf merger scenario can explain the different subclasses of He-sdO by the relative mass fraction contained in the corona as opposed to the accretion disc. Expanding on the work of \citet{Zhang2012}, \citet{Yu2021} found that the masses of the merging white dwarfs also play a role, with lower masses forming N-rich systems and larger masses leading to C-enrichment.
As shown in Fig.~\ref{fig:hrd}, the models of \citet{Yu2021} seem to be able to explain the observed \teff\ and luminosity of the magnetic He-sdOs. However the exact type of merger cannot be constrained, since we cannot place good constraints on C-enrichment, though N-rich surfaces seem to be a characteristic of the four magnetic iHe-sdOs.

Another puzzle is the division of He-sdOs according to hydrogen content into iHe- and eHe-sdOs as discussed extensively by \citet{Luo2021}. All four magnetic He-sdOs show a higher hydrogen abundance than typically observed for He-sdOs \citep[see e.g.][]{Stroeer2007, Schindewolf2018}. However, neither \citet{Yu2021} nor \citet{Justham2011} have included hydrogen in their models. Model predictions are difficult to make, because the atmosphere corresponds to only a small fraction of the stellar envelope. Attempts have been made by \citet{hall2016} and \citet{Schwab2018}, but, as already pointed out by \citet{Dorsch2022}, their models typically predict surfaces poor in hydrogen, at odds with what we find. Yet, we find the stars to lie close to the helium main sequence, which supports that their hydrogen envelopes should be small. The discrepancy between observed and predicted abundances is likely due to limitations on the modelling of the merger, rather than an issue with the idea of a merger itself. For instance, the hydrogen abundance is strongly dependent on rotation, which in turn depends on the angle between the rotation and magnetic axes \citep{GarciaBerro2012}, which is not included in the models. Our fits to the available observations of the magnetic He-sdOs do not constrain the magnetic field geometry well, as that would require higher-resolution spectra allowing to better resolve the shape of the Zeeman components. The fact that more than one homogeneous component was needed to fit the observed spectra already hints at a non-homogeneous magnetic field.

As for the observed projected rotation velocities, they are typically small in hot subdwarfs, irrespective of their chemical composition \citep[see e.g.][]{Geier2012}, and the magnetic systems seem to be no exception, as suggested by our upper limits on \vsini.
As an alternative to a precise $v_\mathrm{rot}\sin i$ estimate that could constrain rotation, we searched for signs of rotation in publicly available TESS, ZTF, and CRTS light curves for the three stars. However, we find no evidence for periodic variability in any of them. Similarly, the magnetic He-sdO from \citet{Dorsch2022} was also found to show no signs of a rotation period in the light curve. Although magnetism is certainly able to induce stellar spots, it seems that detectable spots are uncommon in the case of strongly magnetic He-sdOs.

Apart from mergers, another scenario that could cause magnetism during the hot subdwarf phase is a dynamo acting in the convective core during the main sequence, which has been invoked to explain a fraction of white dwarfs. In this scenario, the field would be exposed when the progenitor star loses its outer layers due to binary interaction. It cannot, however, explain the four known stars given the lack of binary companions. A fossil field from the formation cloud could work similarly, requiring the strongly magnetic Ap and Bp stars to have their cores exposed by binary interaction. The fact that no binary hot subdwarfs have been found to be magnetic could be an argument against these scenarios. The fields in the cores of red giant stars are found to be of the order of $\approx 100$~kG \citep{Fuller2015}, which should be detectable with spectropolarimetry or high-resolution, high signal-to-noise ratio spectra. Only a few tens of hot subdwarfs have spectropolarimetric observations, so the lack of detection in this case is perhaps not surprising. On the other hand, high-resolution spectra are available for hundreds of hot subdwarfs, in particular sdBs. To explain the lack of detection, the fraction of systems with detectable magnetic fields must be a few percent at most, which was also the conclusion of \citet{Landstreet2012}.

\section{Summary \& Conclusions}

We identified three new magnetic hot subdwarfs from their SDSS spectra. Using archival WHT/ISIS spectra and SED fits, we estimated their stellar parameters. The observed magnetic fields are in the range $300-500$~kG. Assuming conservation of magnetic flux, this implies fields of the order of $50-150$~MG at the white dwarf stage, consistent with typically observed values \citep{Kepler2013, Bagnulo2021}. The similarity between the stellar parameters of all four known magnetic hot subdwarfs points at a common origin for all of them. Their lack of radial velocity variability and observed abundances are consistent with a merger channel, though better data, as well as more complete merger models including hydrogen and magnetic fields, are required to constrain the exact channel. In addition, it seems that a merger alone is not sufficient to trigger a magnetic field, given the lack of detection in high-resolution spectra of likely merger remnants, for example by \citet{napi04} and \citet{Werner2022}. Still, our findings provide evidence that mergers are indeed responsible for a fraction of magnetic white dwarfs, in particular those with strong ($\gtrsim 50$~MG) fields.

Formation scenarios other than mergers could lead to magnetism in hot subdwarfs, in particular the stripping of a red giant with a field generated during the main sequence, e.g. due to a convective core. Since evidence of magnetic fields has been found for intermediate-mass red giants \citep[$M \gtrsim 1.1$~M$_{\odot}$,][]{Stello2016}, and those can lead to hot subdwarfs with non-canonical masses (i.e. different from the typical $0.47$~M$_{\sun}$ value resulting from solar-metallicity objects that experience a He-flash), focusing future spectropolarimetric searches on low- or high-mass hot subdwarfs could be profitable. It is worth noting that the stellar-stripping scenario could lead to magnetism also in sdBs -- it predicts He-sdOs that are more luminous than the ones observed here, and sdBs that can have similar luminosities but cooler temperature \citep{Gotberg2018}.

Finally, we propose that an `H' should be added to the spectral class of magnetic hot subdwarfs showing Zeeman splitting, in analogy to white dwarf classes, making \jofour, \jthirteen, \jsixteen, and the prototype \joeight\ from \citet{Dorsch2022} He-sdO{\bf H}s.

\section*{Acknowledgements}

IP and BTG acknowledge support from the UK's Science and Technology Facilities Council (STFC), grant ST/T000406/1. MD and UH acknowledge support by the Deutsche Forschungsgemeinschaft under grant HE1356/70-1. PN acknowledges support from the Grant Agency of the Czech Republic (GA\v{C}R 22-34467S). TK acknowledges support from the National Science Foundation through grant AST \#2107982, from NASA through grant 80NSSC22K0338 and from STScI through grant HST-GO-16659.002-A. Financial support from the National Science Centre under projects No.\,UMO-2017/26/E/ST9/00703 and UMO-2017/25/B/ST9/02218 is acknowledged.
We thank Tom Marsh for the use of the {\sc pamela} and {\sc molly} packages, and Xianfei Zhang for providing us with his most recent evolutionary models.

This project has received funding from the European Research Council (ERC) under the European Union’s Horizon 2020 research and innovation programme (Grant agreement No. 101020057), and from the European Community's Seventh Framework Programme (FP7/2013-2016) under grant agreement number 312430 (OPTICON). Based on observations made under OPTICON 15A/22.


\section*{Data Availability}

All the data analysed in this work are public in the respective data archives.



\bibliographystyle{mnras}
\bibliography{magSDs} 

\begin{thebibliography}{}
\makeatletter
\relax
\def\mn@urlcharsother{\let\do\@makeother \do\$\do\&\do\#\do\^\do\_\do\%\do\~}
\def\mn@doi{\begingroup\mn@urlcharsother \@ifnextchar [ {\mn@doi@}
  {\mn@doi@[]}}
\def\mn@doi@[#1]#2{\def\@tempa{#1}\ifx\@tempa\@empty \href
  {http://dx.doi.org/#2} {doi:#2}\else \href {http://dx.doi.org/#2} {#1}\fi
  \endgroup}
\def\mn@eprint#1#2{\mn@eprint@#1:#2::\@nil}
\def\mn@eprint@arXiv#1{\href {http://arxiv.org/abs/#1} {{\tt arXiv:#1}}}
\def\mn@eprint@dblp#1{\href {http://dblp.uni-trier.de/rec/bibtex/#1.xml}
  {dblp:#1}}
\def\mn@eprint@#1:#2:#3:#4\@nil{\def\@tempa {#1}\def\@tempb {#2}\def\@tempc
  {#3}\ifx \@tempc \@empty \let \@tempc \@tempb \let \@tempb \@tempa \fi \ifx
  \@tempb \@empty \def\@tempb {arXiv}\fi \@ifundefined
  {mn@eprint@\@tempb}{\@tempb:\@tempc}{\expandafter \expandafter \csname
  mn@eprint@\@tempb\endcsname \expandafter{\@tempc}}}

\bibitem[\protect\citeauthoryear{{Ahmad}, {Behara}, {Jeffery}, {Sahin}  \&
  {Woolf}}{{Ahmad} et~al.}{2007}]{Ahmad2007}
{Ahmad} A.,  {Behara} N.~T.,  {Jeffery} C.~S.,  {Sahin} T.,   {Woolf} V.~M.,
  2007, \mn@doi [\aap] {10.1051/0004-6361:20066360}, \href
  {https://ui.adsabs.harvard.edu/abs/2007A&A...465..541A} {465, 541}

\bibitem[\protect\citeauthoryear{{Alam} et~al.,}{{Alam}
  et~al.}{2015}]{Alam2015}
{Alam} S.,  et~al., 2015, \mn@doi [\apjs] {10.1088/0067-0049/219/1/12}, \href
  {https://ui.adsabs.harvard.edu/abs/2015ApJS..219...12A} {219, 12}

\bibitem[\protect\citeauthoryear{{Aller}, {Lillo-Box}, {Jones}, {Miranda}  \&
  {Barcel{\'o} Forteza}}{{Aller} et~al.}{2020}]{tpfplotter}
{Aller} A.,  {Lillo-Box} J.,  {Jones} D.,  {Miranda} L.~F.,   {Barcel{\'o}
  Forteza} S.,  2020, \mn@doi [\aap] {10.1051/0004-6361/201937118}, \href
  {https://ui.adsabs.harvard.edu/abs/2020A&A...635A.128A} {635, A128}

\bibitem[\protect\citeauthoryear{{Angel}, {Borra}  \& {Landstreet}}{{Angel}
  et~al.}{1981}]{Angel1981}
{Angel} J.~R.~P.,  {Borra} E.~F.,   {Landstreet} J.~D.,  1981, \mn@doi [\apjs]
  {10.1086/190720}, \href
  {https://ui.adsabs.harvard.edu/abs/1981ApJS...45..457A} {45, 457}

\bibitem[\protect\citeauthoryear{{Babcock}}{{Babcock}}{1947}]{Babcock1947}
{Babcock} H.~W.,  1947, \mn@doi [\apj] {10.1086/144887}, \href
  {https://ui.adsabs.harvard.edu/abs/1947ApJ...105..105B} {105, 105}

\bibitem[\protect\citeauthoryear{{Bachulski}, {Baran}, {Jeffery},
  {{\O}stensen}, {Reed}, {Telting}  \& {Kuutma}}{{Bachulski}
  et~al.}{2016}]{Bachulski2016}
{Bachulski} S.,  {Baran} A.~S.,  {Jeffery} C.~S.,  {{\O}stensen} R.~H.,  {Reed}
  M.~D.,  {Telting} J.~H.,   {Kuutma} T.,  2016, \actaa, \href
  {https://ui.adsabs.harvard.edu/abs/2016AcA....66..455B} {66, 455}

\bibitem[\protect\citeauthoryear{{Bagnulo} \& {Landstreet}}{{Bagnulo} \&
  {Landstreet}}{2021}]{Bagnulo2021}
{Bagnulo} S.,  {Landstreet} J.~D.,  2021, \mn@doi [\mnras]
  {10.1093/mnras/stab2046}, \href
  {https://ui.adsabs.harvard.edu/abs/2021MNRAS.507.5902B} {507, 5902}

\bibitem[\protect\citeauthoryear{{Bagnulo}, {Landstreet}, {Fossati}  \&
  {Kochukhov}}{{Bagnulo} et~al.}{2012}]{Bagnulo2012}
{Bagnulo} S.,  {Landstreet} J.~D.,  {Fossati} L.,   {Kochukhov} O.,  2012,
  \mn@doi [\aap] {10.1051/0004-6361/201118098}, \href
  {https://ui.adsabs.harvard.edu/abs/2012A&A...538A.129B} {538, A129}

\bibitem[\protect\citeauthoryear{{Bagnulo}, {Fossati}, {Landstreet}  \&
  {Izzo}}{{Bagnulo} et~al.}{2015}]{Bagnulo2015}
{Bagnulo} S.,  {Fossati} L.,  {Landstreet} J.~D.,   {Izzo} C.,  2015, \mn@doi
  [\aap] {10.1051/0004-6361/201526497}, \href
  {https://ui.adsabs.harvard.edu/abs/2015A&A...583A.115B} {583, A115}

\bibitem[\protect\citeauthoryear{{Balona} et~al.,}{{Balona}
  et~al.}{2019}]{Balona2019}
{Balona} L.~A.,  et~al., 2019, \mn@doi [\mnras] {10.1093/mnras/stz586}, \href
  {https://ui.adsabs.harvard.edu/abs/2019MNRAS.485.3457B} {485, 3457}

\bibitem[\protect\citeauthoryear{{Bellm} et~al.,}{{Bellm} et~al.}{2019}]{ztf}
{Bellm} E.~C.,  et~al., 2019, \mn@doi [\pasp] {10.1088/1538-3873/aaecbe}, \href
  {https://ui.adsabs.harvard.edu/abs/2019PASP..131a8002B} {131, 018002}

\bibitem[\protect\citeauthoryear{{Bianchi}, {Shiao}  \& {Thilker}}{{Bianchi}
  et~al.}{2017}]{Bianchi2017}
{Bianchi} L.,  {Shiao} B.,   {Thilker} D.,  2017, \mn@doi [\apjs]
  {10.3847/1538-4365/aa7053}, \href
  {https://ui.adsabs.harvard.edu/abs/2017ApJS..230...24B} {230, 24}

\bibitem[\protect\citeauthoryear{{Briggs}, {Ferrario}, {Tout}, {Wickramasinghe}
   \& {Hurley}}{{Briggs} et~al.}{2015}]{Briggs2015}
{Briggs} G.~P.,  {Ferrario} L.,  {Tout} C.~A.,  {Wickramasinghe} D.~T.,
  {Hurley} J.~R.,  2015, \mn@doi [\mnras] {10.1093/mnras/stu2539}, \href
  {https://ui.adsabs.harvard.edu/abs/2015MNRAS.447.1713B} {447, 1713}

\bibitem[\protect\citeauthoryear{{Briggs}, {Ferrario}, {Tout}  \&
  {Wickramasinghe}}{{Briggs} et~al.}{2018}]{Briggs2018}
{Briggs} G.~P.,  {Ferrario} L.,  {Tout} C.~A.,   {Wickramasinghe} D.~T.,  2018,
  \mn@doi [\mnras] {10.1093/mnras/sty1150}, \href
  {https://ui.adsabs.harvard.edu/abs/2018MNRAS.478..899B} {478, 899}

\bibitem[\protect\citeauthoryear{{Charpinet}, {Giammichele}, {Zong}, {Van
  Grootel}, {Brassard}  \& {Fontaine}}{{Charpinet}
  et~al.}{2018}]{Charpinet2018}
{Charpinet} S.,  {Giammichele} N.,  {Zong} W.,  {Van Grootel} V.,  {Brassard}
  P.,   {Fontaine} G.,  2018, \mn@doi [Open Astronomy]
  {10.1515/astro-2018-0012}, \href
  {https://ui.adsabs.harvard.edu/abs/2018OAst...27..112C} {27, 112}

\bibitem[\protect\citeauthoryear{{Chountonov} \& {Geier}}{{Chountonov} \&
  {Geier}}{2012}]{Chountonov2012}
{Chountonov} G.,  {Geier} S.,  2012, in {Kilkenny} D.,  {Jeffery} C.~S.,
  {Koen} C.,  eds,  Astronomical Society of the Pacific Conference Series Vol.
  452, Fifth Meeting on Hot Subdwarf Stars and Related Objects. p.~93
  (\mn@eprint {arXiv} {1112.2921})

\bibitem[\protect\citeauthoryear{{Copperwheat}, {Morales-Rueda}, {Marsh},
  {Maxted}  \& {Heber}}{{Copperwheat} et~al.}{2011}]{Copperwheat2011}
{Copperwheat} C.~M.,  {Morales-Rueda} L.,  {Marsh} T.~R.,  {Maxted} P.~F.~L.,
  {Heber} U.,  2011, \mn@doi [\mnras] {10.1111/j.1365-2966.2011.18786.x}, \href
  {https://ui.adsabs.harvard.edu/abs/2011MNRAS.415.1381C} {415, 1381}

\bibitem[\protect\citeauthoryear{{Culpan}, {Geier}, {Reindl}, {Pelisoli},
  {Gentile-Fusillo}  \& {Vorontseva}}{{Culpan} et~al.}{2022}]{Culpan2022}
{Culpan} R.,  {Geier} S.,  {Reindl} N.,  {Pelisoli} I.,  {Gentile-Fusillo} N.,
   {Vorontseva} A.,  2022, \aap, in press

\bibitem[\protect\citeauthoryear{{Cutri} et~al.,}{{Cutri}
  et~al.}{2003}]{Cutri2003}
{Cutri} R.~M.,  et~al., 2003, VizieR Online Data Catalog, \href
  {https://ui.adsabs.harvard.edu/abs/2003yCat.2246....0C} {p. II/246}

\bibitem[\protect\citeauthoryear{{Dorman}, {Rood}  \& {O'Connell}}{{Dorman}
  et~al.}{1993}]{Dorman1993}
{Dorman} B.,  {Rood} R.~T.,   {O'Connell} R.~W.,  1993, \mn@doi [\apj]
  {10.1086/173511}, \href
  {https://ui.adsabs.harvard.edu/abs/1993ApJ...419..596D} {419, 596}

\bibitem[\protect\citeauthoryear{{Dorsch}}{{Dorsch}}{2022}]{Dorsch2022b}
{Dorsch} M.,  2022, PhD thesis, Friedrich-Alexander University
  Erlangen-N{\"u}rnberg, in prep.

\bibitem[\protect\citeauthoryear{{Dorsch}, {Latour}  \& {Heber}}{{Dorsch}
  et~al.}{2019}]{Dorsch2019}
{Dorsch} M.,  {Latour} M.,   {Heber} U.,  2019, \mn@doi [\aap]
  {10.1051/0004-6361/201935724}, \href
  {https://ui.adsabs.harvard.edu/abs/2019A&A...630A.130D} {630, A130}

\bibitem[\protect\citeauthoryear{{Dorsch}, {Latour}, {Heber}, {Irrgang},
  {Charpinet}  \& {Jeffery}}{{Dorsch} et~al.}{2020}]{Dorsch2020}
{Dorsch} M.,  {Latour} M.,  {Heber} U.,  {Irrgang} A.,  {Charpinet} S.,
  {Jeffery} C.~S.,  2020, \mn@doi [\aap] {10.1051/0004-6361/202038859}, \href
  {https://ui.adsabs.harvard.edu/abs/2020A&A...643A..22D} {643, A22}

\bibitem[\protect\citeauthoryear{{Dorsch}, {Reindl}, {Pelisoli}, {Heber},
  {Geier}, {Istrate}  \& {Justham}}{{Dorsch} et~al.}{2022}]{Dorsch2022}
{Dorsch} M.,  {Reindl} N.,  {Pelisoli} I.,  {Heber} U.,  {Geier} S.,  {Istrate}
  A.~G.,   {Justham} S.,  2022, arXiv e-prints, \href
  {https://ui.adsabs.harvard.edu/abs/2022arXiv220108146D} {p. arXiv:2201.08146}

\bibitem[\protect\citeauthoryear{{Drake} et~al.,}{{Drake} et~al.}{2009}]{crts}
{Drake} A.~J.,  et~al., 2009, \mn@doi [\apj] {10.1088/0004-637X/696/1/870},
  \href {https://ui.adsabs.harvard.edu/abs/2009ApJ...696..870D} {696, 870}

\bibitem[\protect\citeauthoryear{{Edelmann}, {Heber}, {Altmann}, {Karl}  \&
  {Lisker}}{{Edelmann} et~al.}{2005}]{Edelmann2005}
{Edelmann} H.,  {Heber} U.,  {Altmann} M.,  {Karl} C.,   {Lisker} T.,  2005,
  \mn@doi [\aap] {10.1051/0004-6361:20053267}, \href
  {https://ui.adsabs.harvard.edu/abs/2005A&A...442.1023E} {442, 1023}

\bibitem[\protect\citeauthoryear{{Eisenstein} et~al.,}{{Eisenstein}
  et~al.}{2011}]{sdssiii}
{Eisenstein} D.~J.,  et~al., 2011, \mn@doi [\aj] {10.1088/0004-6256/142/3/72},
  \href {https://ui.adsabs.harvard.edu/abs/2011AJ....142...72E} {142, 72}

\bibitem[\protect\citeauthoryear{{El-Badry}, {Rix}  \& {Heintz}}{{El-Badry}
  et~al.}{2021}]{El-Badry2021}
{El-Badry} K.,  {Rix} H.-W.,   {Heintz} T.~M.,  2021, \mn@doi [\mnras]
  {10.1093/mnras/stab323}, \href
  {https://ui.adsabs.harvard.edu/abs/2021MNRAS.506.2269E} {506, 2269}

\bibitem[\protect\citeauthoryear{{Elkin}}{{Elkin}}{1996}]{Elkin1996}
{Elkin} V.~G.,  1996, \aap, \href
  {https://ui.adsabs.harvard.edu/abs/1996A&A...312L...5E} {312, L5}

\bibitem[\protect\citeauthoryear{{Elkin}}{{Elkin}}{1998}]{Elkin1998}
{Elkin} V.~G.,  1998, Contributions of the Astronomical Observatory Skalnate
  Pleso, \href {https://ui.adsabs.harvard.edu/abs/1998CoSka..27..452E} {27,
  452}

\bibitem[\protect\citeauthoryear{{Feinstein} et~al.,}{{Feinstein}
  et~al.}{2019}]{Feinstein2019}
{Feinstein} A.~D.,  et~al., 2019, \mn@doi [\pasp] {10.1088/1538-3873/ab291c},
  \href {https://ui.adsabs.harvard.edu/abs/2019PASP..131i4502F} {131, 094502}

\bibitem[\protect\citeauthoryear{{Ferrario}, {Pringle}, {Tout}  \&
  {Wickramasinghe}}{{Ferrario} et~al.}{2009}]{Ferrario2009}
{Ferrario} L.,  {Pringle} J.~E.,  {Tout} C.~A.,   {Wickramasinghe} D.~T.,
  2009, \mn@doi [\mnras] {10.1111/j.1745-3933.2009.00765.x}, \href
  {https://ui.adsabs.harvard.edu/abs/2009MNRAS.400L..71F} {400, L71}

\bibitem[\protect\citeauthoryear{{Ferrario}, {de Martino}  \&
  {G{\"a}nsicke}}{{Ferrario} et~al.}{2015}]{Ferrario2015}
{Ferrario} L.,  {de Martino} D.,   {G{\"a}nsicke} B.~T.,  2015, \mn@doi [\ssr]
  {10.1007/s11214-015-0152-0}, \href
  {https://ui.adsabs.harvard.edu/abs/2015SSRv..191..111F} {191, 111}

\bibitem[\protect\citeauthoryear{{Fitzpatrick}, {Massa}, {Gordon}, {Bohlin}  \&
  {Clayton}}{{Fitzpatrick} et~al.}{2019}]{Fitzpatrick2019}
{Fitzpatrick} E.~L.,  {Massa} D.,  {Gordon} K.~D.,  {Bohlin} R.,   {Clayton}
  G.~C.,  2019, \mn@doi [\apj] {10.3847/1538-4357/ab4c3a}, \href
  {https://ui.adsabs.harvard.edu/abs/2019ApJ...886..108F} {886, 108}

\bibitem[\protect\citeauthoryear{{Fuller}, {Cantiello}, {Stello}, {Garcia}  \&
  {Bildsten}}{{Fuller} et~al.}{2015}]{Fuller2015}
{Fuller} J.,  {Cantiello} M.,  {Stello} D.,  {Garcia} R.~A.,   {Bildsten} L.,
  2015, \mn@doi [Science] {10.1126/science.aac6933}, \href
  {https://ui.adsabs.harvard.edu/abs/2015Sci...350..423F} {350, 423}

\bibitem[\protect\citeauthoryear{{Gaia Collaboration} et~al.,}{{Gaia
  Collaboration} et~al.}{2016}]{Gaia2016}
{Gaia Collaboration} et~al., 2016, \mn@doi [\aap]
  {10.1051/0004-6361/201629272}, \href
  {https://ui.adsabs.harvard.edu/abs/2016A&A...595A...1G} {595, A1}

\bibitem[\protect\citeauthoryear{{Gaia Collaboration} et~al.,}{{Gaia
  Collaboration} et~al.}{2021}]{Gaia2020}
{Gaia Collaboration} et~al., 2021, \mn@doi [\aap]
  {10.1051/0004-6361/202039657}, \href
  {https://ui.adsabs.harvard.edu/abs/2021A&A...649A...1G} {649, A1}

\bibitem[\protect\citeauthoryear{{Garc{\'\i}a-Berro}
  et~al.,}{{Garc{\'\i}a-Berro} et~al.}{2012}]{GarciaBerro2012}
{Garc{\'\i}a-Berro} E.,  et~al., 2012, \mn@doi [\apj]
  {10.1088/0004-637X/749/1/25}, \href
  {https://ui.adsabs.harvard.edu/abs/2012ApJ...749...25G} {749, 25}

\bibitem[\protect\citeauthoryear{{Geier}}{{Geier}}{2020}]{Geier2020}
{Geier} S.,  2020, \mn@doi [\aap] {10.1051/0004-6361/202037526}, \href
  {https://ui.adsabs.harvard.edu/abs/2020A&A...635A.193G} {635, A193}

\bibitem[\protect\citeauthoryear{{Geier} \& {Heber}}{{Geier} \&
  {Heber}}{2012}]{Geier2012}
{Geier} S.,  {Heber} U.,  2012, \mn@doi [\aap] {10.1051/0004-6361/201219463},
  \href {https://ui.adsabs.harvard.edu/abs/2012A&A...543A.149G} {543, A149}

\bibitem[\protect\citeauthoryear{{Geier}, {Heber}, {Kupfer}  \&
  {Napiwotzki}}{{Geier} et~al.}{2010a}]{Geier2010spy}
{Geier} S.,  {Heber} U.,  {Kupfer} T.,   {Napiwotzki} R.,  2010a, \mn@doi
  [\aap] {10.1051/0004-6361/200912545}, \href
  {https://ui.adsabs.harvard.edu/abs/2010A&A...515A..37G} {515, A37}

\bibitem[\protect\citeauthoryear{{Geier}, {Heber}, {Podsiadlowski}, {Edelmann},
  {Napiwotzki}, {Kupfer}  \& {M{\"u}ller}}{{Geier} et~al.}{2010b}]{Geier2010}
{Geier} S.,  {Heber} U.,  {Podsiadlowski} P.,  {Edelmann} H.,  {Napiwotzki} R.,
   {Kupfer} T.,   {M{\"u}ller} S.,  2010b, \mn@doi [\aap]
  {10.1051/0004-6361/201014465}, \href
  {https://ui.adsabs.harvard.edu/abs/2010A&A...519A..25G} {519, A25}

\bibitem[\protect\citeauthoryear{{Geier}, {Heber}, {Edelmann}, {Morales-Rueda},
  {Kilkenny}, {O'Donoghue}, {Marsh}  \& {Copperwheat}}{{Geier}
  et~al.}{2013}]{Geier2013}
{Geier} S.,  {Heber} U.,  {Edelmann} H.,  {Morales-Rueda} L.,  {Kilkenny} D.,
  {O'Donoghue} D.,  {Marsh} T.~R.,   {Copperwheat} C.,  2013, \mn@doi [\aap]
  {10.1051/0004-6361/201322057}, \href
  {https://ui.adsabs.harvard.edu/abs/2013A&A...557A.122G} {557, A122}

\bibitem[\protect\citeauthoryear{{Geier} et~al.,}{{Geier}
  et~al.}{2014}]{Geier2014}
{Geier} S.,  et~al., 2014, \mn@doi [\aap] {10.1051/0004-6361/201323115}, \href
  {https://ui.adsabs.harvard.edu/abs/2014A&A...562A..95G} {562, A95}

\bibitem[\protect\citeauthoryear{{Geier} et~al.,}{{Geier}
  et~al.}{2015}]{Geier2015}
{Geier} S.,  et~al., 2015, \mn@doi [\aap] {10.1051/0004-6361/201525666}, \href
  {https://ui.adsabs.harvard.edu/abs/2015A&A...577A..26G} {577, A26}

\bibitem[\protect\citeauthoryear{{Geier}, {Dorsch}, {Pelisoli}, {Reindl},
  {Heber}  \& {Irrgang}}{{Geier} et~al.}{2022}]{Geier2022}
{Geier} S.,  {Dorsch} M.,  {Pelisoli} I.,  {Reindl} N.,  {Heber} U.,
  {Irrgang} A.,  2022, arXiv e-prints, \href
  {https://ui.adsabs.harvard.edu/abs/2022arXiv220209608G} {p. arXiv:2202.09608}

\bibitem[\protect\citeauthoryear{{G{\"o}tberg}, {de Mink}, {Groh}, {Kupfer},
  {Crowther}, {Zapartas}  \& {Renzo}}{{G{\"o}tberg} et~al.}{2018}]{Gotberg2018}
{G{\"o}tberg} Y.,  {de Mink} S.~E.,  {Groh} J.~H.,  {Kupfer} T.,  {Crowther}
  P.~A.,  {Zapartas} E.,   {Renzo} M.,  2018, \mn@doi [\aap]
  {10.1051/0004-6361/201732274}, \href
  {https://ui.adsabs.harvard.edu/abs/2018A&A...615A..78G} {615, A78}

\bibitem[\protect\citeauthoryear{{Hall} \& {Jeffery}}{{Hall} \&
  {Jeffery}}{2016}]{hall2016}
{Hall} P.~D.,  {Jeffery} C.~S.,  2016, \mn@doi [\mnras]
  {10.1093/mnras/stw2188}, \href
  {https://ui.adsabs.harvard.edu/abs/2016MNRAS.463.2756H} {463, 2756}

\bibitem[\protect\citeauthoryear{{Han}, {Podsiadlowski}, {Maxted}, {Marsh}  \&
  {Ivanova}}{{Han} et~al.}{2002}]{Han2002}
{Han} Z.,  {Podsiadlowski} P.,  {Maxted} P.~F.~L.,  {Marsh} T.~R.,   {Ivanova}
  N.,  2002, \mn@doi [\mnras] {10.1046/j.1365-8711.2002.05752.x}, \href
  {https://ui.adsabs.harvard.edu/abs/2002MNRAS.336..449H} {336, 449}

\bibitem[\protect\citeauthoryear{{Han}, {Podsiadlowski}, {Maxted}  \&
  {Marsh}}{{Han} et~al.}{2003}]{Han2003}
{Han} Z.,  {Podsiadlowski} P.,  {Maxted} P.~F.~L.,   {Marsh} T.~R.,  2003,
  \mn@doi [\mnras] {10.1046/j.1365-8711.2003.06451.x}, \href
  {https://ui.adsabs.harvard.edu/abs/2003MNRAS.341..669H} {341, 669}

\bibitem[\protect\citeauthoryear{{Heber}}{{Heber}}{2016}]{Heber2016}
{Heber} U.,  2016, \mn@doi [\pasp] {10.1088/1538-3873/128/966/082001}, \href
  {https://ui.adsabs.harvard.edu/abs/2016PASP..128h2001H} {128, 082001}

\bibitem[\protect\citeauthoryear{{Heber} \& {Hunger}}{{Heber} \&
  {Hunger}}{1987}]{Heber1987}
{Heber} U.,  {Hunger} K.,  1987, The Messenger, \href
  {https://ui.adsabs.harvard.edu/abs/1987Msngr..47...36H} {47, 36}

\bibitem[\protect\citeauthoryear{{Heber}, {Edelmann}, {Lisker}  \&
  {Napiwotzki}}{{Heber} et~al.}{2003}]{Heber2003}
{Heber} U.,  {Edelmann} H.,  {Lisker} T.,   {Napiwotzki} R.,  2003, \mn@doi
  [\aap] {10.1051/0004-6361:20031553}, \href
  {https://ui.adsabs.harvard.edu/abs/2003A&A...411L.477H} {411, L477}

\bibitem[\protect\citeauthoryear{{Heber}, {Geier}  \& {Gaensicke}}{{Heber}
  et~al.}{2013}]{Heber2013}
{Heber} U.,  {Geier} S.,   {Gaensicke} B.,  2013, in European Physical Journal
  Web of Conferences. p. 04002 (\mn@eprint {arXiv} {1211.5315}),
  \mn@doi{10.1051/epjconf/20134304002}

\bibitem[\protect\citeauthoryear{{Henden}, {Templeton}, {Terrell}, {Smith},
  {Levine}  \& {Welch}}{{Henden} et~al.}{2016}]{Henden2016}
{Henden} A.~A.,  {Templeton} M.,  {Terrell} D.,  {Smith} T.~C.,  {Levine} S.,
  {Welch} D.,  2016, VizieR Online Data Catalog, \href
  {https://ui.adsabs.harvard.edu/abs/2016yCat.2336....0H} {p. II/336}

\bibitem[\protect\citeauthoryear{{Herbig}}{{Herbig}}{1999}]{Herbig1999}
{Herbig} G.~H.,  1999, \mn@doi [\pasp] {10.1086/316426}, \href
  {https://ui.adsabs.harvard.edu/abs/1999PASP..111.1144H} {111, 1144}

\bibitem[\protect\citeauthoryear{{Hirsch}}{{Hirsch}}{2009}]{hirsch09}
{Hirsch} H.~A.,  2009, PhD thesis, Friedrich-Alexander University
  Erlangen-N{\"u}rnberg

\bibitem[\protect\citeauthoryear{{Howarth} \& {Heber}}{{Howarth} \&
  {Heber}}{1990}]{Howarth1990}
{Howarth} I.~D.,  {Heber} U.,  1990, \mn@doi [\pasp] {10.1086/132716}, \href
  {https://ui.adsabs.harvard.edu/abs/1990PASP..102..912H} {102, 912}

\bibitem[\protect\citeauthoryear{{Hubeny} \& {Lanz}}{{Hubeny} \&
  {Lanz}}{2017c}]{hubeny17a}
{Hubeny} I.,  {Lanz} T.,  2017c, preprint, \href
  {http://adsabs.harvard.edu/abs/2017arXiv170601859H} {} (\mn@eprint {arXiv}
  {1706.01859})

\bibitem[\protect\citeauthoryear{{Hubeny} \& {Lanz}}{{Hubeny} \&
  {Lanz}}{2017a}]{hubeny17b}
{Hubeny} I.,  {Lanz} T.,  2017a, preprint, \href
  {http://adsabs.harvard.edu/abs/2017arXiv170601935H} {} (\mn@eprint {arXiv}
  {1706.01935})

\bibitem[\protect\citeauthoryear{{Hubeny} \& {Lanz}}{{Hubeny} \&
  {Lanz}}{2017b}]{hubeny17c}
{Hubeny} I.,  {Lanz} T.,  2017b, preprint, \href
  {http://adsabs.harvard.edu/abs/2017arXiv170601937H} {} (\mn@eprint {arXiv}
  {1706.01937})

\bibitem[\protect\citeauthoryear{{Husfeld}, {Butler}, {Heber}  \&
  {Drilling}}{{Husfeld} et~al.}{1989}]{Husfeld1989}
{Husfeld} D.,  {Butler} K.,  {Heber} U.,   {Drilling} J.~S.,  1989, \aap, \href
  {https://ui.adsabs.harvard.edu/abs/1989A&A...222..150H} {222, 150}

\bibitem[\protect\citeauthoryear{{Isern}, {Garc{\'\i}a-Berro}, {K{\"u}lebi}  \&
  {Lor{\'e}n-Aguilar}}{{Isern} et~al.}{2017}]{Isern2017}
{Isern} J.,  {Garc{\'\i}a-Berro} E.,  {K{\"u}lebi} B.,   {Lor{\'e}n-Aguilar}
  P.,  2017, \mn@doi [\apjl] {10.3847/2041-8213/aa5eae}, \href
  {https://ui.adsabs.harvard.edu/abs/2017ApJ...836L..28I} {836, L28}

\bibitem[\protect\citeauthoryear{{Jacobs} et~al.,}{{Jacobs}
  et~al.}{2011}]{Jacobs2011}
{Jacobs} V.~A.,  et~al., 2011, in {Schuh} S.,  {Drechsel} H.,   {Heber} U.,
  eds,  American Institute of Physics Conference Series Vol. 1331, Planetary
  Systems Beyond the Main Sequence. pp 304--309 (\mn@eprint {arXiv}
  {1101.4158}), \mn@doi{10.1063/1.3556216}

\bibitem[\protect\citeauthoryear{{Jeffery} et~al.,}{{Jeffery}
  et~al.}{2013}]{Jeffery2013}
{Jeffery} C.~S.,  et~al., 2013, \mn@doi [\mnras] {10.1093/mnras/sts579}, \href
  {https://ui.adsabs.harvard.edu/abs/2013MNRAS.429.3207J} {429, 3207}

\bibitem[\protect\citeauthoryear{{Jeffery}, {Ahmad}, {Naslim}  \&
  {Kerzendorf}}{{Jeffery} et~al.}{2015}]{Jeffery2015}
{Jeffery} C.~S.,  {Ahmad} A.,  {Naslim} N.,   {Kerzendorf} W.,  2015, \mn@doi
  [\mnras] {10.1093/mnras/stu2203}, \href
  {https://ui.adsabs.harvard.edu/abs/2015MNRAS.446.1889J} {446, 1889}

\bibitem[\protect\citeauthoryear{{Jeffery}, {Miszalski}  \&
  {Snowdon}}{{Jeffery} et~al.}{2021}]{Jeffery2021}
{Jeffery} C.~S.,  {Miszalski} B.,   {Snowdon} E.,  2021, \mn@doi [\mnras]
  {10.1093/mnras/staa3648}, \href
  {https://ui.adsabs.harvard.edu/abs/2021MNRAS.501..623J} {501, 623}

\bibitem[\protect\citeauthoryear{{Justham}, {Podsiadlowski}  \&
  {Han}}{{Justham} et~al.}{2011}]{Justham2011}
{Justham} S.,  {Podsiadlowski} P.,   {Han} Z.,  2011, \mn@doi [\mnras]
  {10.1111/j.1365-2966.2010.17497.x}, \href
  {https://ui.adsabs.harvard.edu/abs/2011MNRAS.410..984J} {410, 984}

\bibitem[\protect\citeauthoryear{{Kawka}, {Vennes}, {Schmidt}, {Wickramasinghe}
   \& {Koch}}{{Kawka} et~al.}{2007}]{Kawka2007}
{Kawka} A.,  {Vennes} S.,  {Schmidt} G.~D.,  {Wickramasinghe} D.~T.,   {Koch}
  R.,  2007, \mn@doi [\apj] {10.1086/509072}, \href
  {https://ui.adsabs.harvard.edu/abs/2007ApJ...654..499K} {654, 499}

\bibitem[\protect\citeauthoryear{{Kawka}, {Vennes}, {O'Toole}, {N{\'e}meth},
  {Burton}, {Kotze}  \& {Buckley}}{{Kawka} et~al.}{2015}]{Kawka2015}
{Kawka} A.,  {Vennes} S.,  {O'Toole} S.,  {N{\'e}meth} P.,  {Burton} D.,
  {Kotze} E.,   {Buckley} D.~A.~H.,  2015, \mn@doi [\mnras]
  {10.1093/mnras/stv821}, \href
  {https://ui.adsabs.harvard.edu/abs/2015MNRAS.450.3514K} {450, 3514}

\bibitem[\protect\citeauthoryear{{Kemp}, {Swedlund}, {Landstreet}  \&
  {Angel}}{{Kemp} et~al.}{1970}]{Kemp1970}
{Kemp} J.~C.,  {Swedlund} J.~B.,  {Landstreet} J.~D.,   {Angel} J.~R.~P.,
  1970, \mn@doi [\apjl] {10.1086/180574}, \href
  {https://ui.adsabs.harvard.edu/abs/1970ApJ...161L..77K} {161, L77}

\bibitem[\protect\citeauthoryear{{Kepler} et~al.,}{{Kepler}
  et~al.}{2013}]{Kepler2013}
{Kepler} S.~O.,  et~al., 2013, \mn@doi [\mnras] {10.1093/mnras/sts522}, \href
  {https://ui.adsabs.harvard.edu/abs/2013MNRAS.429.2934K} {429, 2934}

\bibitem[\protect\citeauthoryear{{Khalack}, {Yameogo}, {LeBlanc}, {Fontaine},
  {Green}, {Van Grootel}  \& {Petit}}{{Khalack} et~al.}{2014}]{Khalack2014}
{Khalack} V.,  {Yameogo} B.,  {LeBlanc} F.,  {Fontaine} G.,  {Green} E.,  {Van
  Grootel} V.,   {Petit} P.,  2014, \mn@doi [\mnras] {10.1093/mnras/stu2012},
  \href {https://ui.adsabs.harvard.edu/abs/2014MNRAS.445.4086K} {445, 4086}

\bibitem[\protect\citeauthoryear{{Kilkenny}, {Heber}  \& {Drilling}}{{Kilkenny}
  et~al.}{1988}]{Kilkenny1988}
{Kilkenny} D.,  {Heber} U.,   {Drilling} J.~S.,  1988, South African
  Astronomical Observatory Circular, \href
  {https://ui.adsabs.harvard.edu/abs/1988SAAOC..12....1K} {12, 1}

\bibitem[\protect\citeauthoryear{{Lallement} et~al.,}{{Lallement}
  et~al.}{2018}]{Lallement2018}
{Lallement} R.,  et~al., 2018, \mn@doi [\aap] {10.1051/0004-6361/201832832},
  \href {https://ui.adsabs.harvard.edu/abs/2018A&A...616A.132L} {616, A132}

\bibitem[\protect\citeauthoryear{{Landstreet}}{{Landstreet}}{1967}]{Landstreet1967}
{Landstreet} J.~D.,  1967, \mn@doi [Physical Review]
  {10.1103/PhysRev.153.1372}, \href
  {https://ui.adsabs.harvard.edu/abs/1967PhRv..153.1372L} {153, 1372}

\bibitem[\protect\citeauthoryear{{Landstreet}, {Bagnulo}, {Fossati}, {Jordan}
  \& {O'Toole}}{{Landstreet} et~al.}{2012}]{Landstreet2012}
{Landstreet} J.~D.,  {Bagnulo} S.,  {Fossati} L.,  {Jordan} S.,   {O'Toole}
  S.~J.,  2012, \mn@doi [\aap] {10.1051/0004-6361/201219178}, \href
  {https://ui.adsabs.harvard.edu/abs/2012A&A...541A.100L} {541, A100}

\bibitem[\protect\citeauthoryear{{Lanz}, {Hubeny}  \& {Heap}}{{Lanz}
  et~al.}{1997}]{Lanz1997}
{Lanz} T.,  {Hubeny} I.,   {Heap} S.~R.,  1997, \mn@doi [\apj]
  {10.1086/304455}, \href
  {https://ui.adsabs.harvard.edu/abs/1997ApJ...485..843L} {485, 843}

\bibitem[\protect\citeauthoryear{{Latour}, {Fontaine}, {Green}  \&
  {Brassard}}{{Latour} et~al.}{2015}]{Latour2015}
{Latour} M.,  {Fontaine} G.,  {Green} E.~M.,   {Brassard} P.,  2015, \mn@doi
  [\aap] {10.1051/0004-6361/201525999}, \href
  {https://ui.adsabs.harvard.edu/abs/2015A&A...579A..39L} {579, A39}

\bibitem[\protect\citeauthoryear{{Latour}, {Chayer}, {Green}, {Irrgang}  \&
  {Fontaine}}{{Latour} et~al.}{2018}]{Latour2018}
{Latour} M.,  {Chayer} P.,  {Green} E.~M.,  {Irrgang} A.,   {Fontaine} G.,
  2018, \mn@doi [\aap] {10.1051/0004-6361/201731496}, \href
  {https://ui.adsabs.harvard.edu/abs/2018A&A...609A..89L} {609, A89}

\bibitem[\protect\citeauthoryear{{Lawrence} et~al.,}{{Lawrence}
  et~al.}{2007}]{ukidss}
{Lawrence} A.,  et~al., 2007, \mn@doi [\mnras]
  {10.1111/j.1365-2966.2007.12040.x}, \href
  {https://ui.adsabs.harvard.edu/abs/2007MNRAS.379.1599L} {379, 1599}

\bibitem[\protect\citeauthoryear{{Lei}, {Zhao}, {N{\'e}meth}  \& {Zhao}}{{Lei}
  et~al.}{2018}]{Lei2018}
{Lei} Z.,  {Zhao} J.,  {N{\'e}meth} P.,   {Zhao} G.,  2018, \mn@doi [\apj]
  {10.3847/1538-4357/aae82b}, \href
  {https://ui.adsabs.harvard.edu/abs/2018ApJ...868...70L} {868, 70}

\bibitem[\protect\citeauthoryear{{Lindegren} et~al.,}{{Lindegren}
  et~al.}{2021}]{Lindegren2021}
{Lindegren} L.,  et~al., 2021, \mn@doi [\aap] {10.1051/0004-6361/202039653},
  \href {https://ui.adsabs.harvard.edu/abs/2021A&A...649A...4L} {649, A4}

\bibitem[\protect\citeauthoryear{{Lisker}, {Heber}, {Napiwotzki}, {Christlieb},
  {Han}, {Homeier}  \& {Reimers}}{{Lisker} et~al.}{2005}]{Lisker2005}
{Lisker} T.,  {Heber} U.,  {Napiwotzki} R.,  {Christlieb} N.,  {Han} Z.,
  {Homeier} D.,   {Reimers} D.,  2005, \mn@doi [\aap]
  {10.1051/0004-6361:20040232}, \href
  {https://ui.adsabs.harvard.edu/abs/2005A&A...430..223L} {430, 223}

\bibitem[\protect\citeauthoryear{{Luo}, {N{\'e}meth}, {Wang}, {Wang}  \&
  {Han}}{{Luo} et~al.}{2021}]{Luo2021}
{Luo} Y.,  {N{\'e}meth} P.,  {Wang} K.,  {Wang} X.,   {Han} Z.,  2021, \mn@doi
  [\apjs] {10.3847/1538-4365/ac11f6}, \href
  {https://ui.adsabs.harvard.edu/abs/2021ApJS..256...28L} {256, 28}

\bibitem[\protect\citeauthoryear{{Magnier} et~al.,}{{Magnier}
  et~al.}{2020}]{Magnier2020}
{Magnier} E.~A.,  et~al., 2020, \mn@doi [\apjs] {10.3847/1538-4365/abb82a},
  \href {https://ui.adsabs.harvard.edu/abs/2020ApJS..251....6M} {251, 6}

\bibitem[\protect\citeauthoryear{{Marsh}}{{Marsh}}{1989}]{Marsh1989}
{Marsh} T.~R.,  1989, \mn@doi [\pasp] {10.1086/132570}, \href
  {https://ui.adsabs.harvard.edu/abs/1989PASP..101.1032M} {101, 1032}

\bibitem[\protect\citeauthoryear{{Mathys}, {Hubrig}, {Mason}, {Michaud},
  {Sch{\"o}ller}  \& {Wesemael}}{{Mathys} et~al.}{2012}]{Mathys2012}
{Mathys} G.,  {Hubrig} S.,  {Mason} E.,  {Michaud} G.,  {Sch{\"o}ller} M.,
  {Wesemael} F.,  2012, \mn@doi [Astronomische Nachrichten]
  {10.1002/asna.201111618}, \href
  {https://ui.adsabs.harvard.edu/abs/2012AN....333...30M} {333, 30}

\bibitem[\protect\citeauthoryear{{Maxted}, {Heber}, {Marsh}  \&
  {North}}{{Maxted} et~al.}{2001}]{maxted01}
{Maxted} P.~F.~L.,  {Heber} U.,  {Marsh} T.~R.,   {North} R.~C.,  2001, \mn@doi
  [\mnras] {10.1111/j.1365-2966.2001.04714.x}, \href
  {https://ui.adsabs.harvard.edu/abs/2001MNRAS.326.1391M} {326, 1391}

\bibitem[\protect\citeauthoryear{{Miller Bertolami}}{{Miller
  Bertolami}}{2016}]{MillerBertolami2016}
{Miller Bertolami} M.~M.,  2016, \mn@doi [\aap] {10.1051/0004-6361/201526577},
  \href {https://ui.adsabs.harvard.edu/abs/2016A&A...588A..25M} {588, A25}

\bibitem[\protect\citeauthoryear{{Miller Bertolami}, {Battich}, {C{\'o}rsico},
  {Althaus}  \& {Wachlin}}{{Miller Bertolami}
  et~al.}{2022}]{MillerBertolami2022}
{Miller Bertolami} M.~M.,  {Battich} T.,  {C{\'o}rsico} A.~H.,  {Althaus}
  L.~G.,   {Wachlin} F.~C.,  2022, arXiv e-prints, \href
  {https://ui.adsabs.harvard.edu/abs/2022arXiv220205635M} {p. arXiv:2202.05635}

\bibitem[\protect\citeauthoryear{{Momany} et~al.,}{{Momany}
  et~al.}{2020}]{Momany2020}
{Momany} Y.,  et~al., 2020, \mn@doi [Nature Astronomy]
  {10.1038/s41550-020-1113-4}, \href
  {https://ui.adsabs.harvard.edu/abs/2020NatAs...4.1092M} {4, 1092}

\bibitem[\protect\citeauthoryear{{Moss}}{{Moss}}{2001}]{Moss2001}
{Moss} D.,  2001, in {Mathys} G.,  {Solanki} S.~K.,   {Wickramasinghe} D.~T.,
  eds,  Astronomical Society of the Pacific Conference Series Vol. 248,
  Magnetic Fields Across the Hertzsprung-Russell Diagram. p.~305

\bibitem[\protect\citeauthoryear{{Napiwotzki} et~al.,}{{Napiwotzki}
  et~al.}{2003}]{napi03}
{Napiwotzki} R.,  et~al., 2003, The Messenger, \href
  {https://ui.adsabs.harvard.edu/abs/2003Msngr.112...25N} {112, 25}

\bibitem[\protect\citeauthoryear{{Napiwotzki}, {Karl}, {Lisker}, {Heber},
  {Christlieb}, {Reimers}, {Nelemans}  \& {Homeier}}{{Napiwotzki}
  et~al.}{2004}]{napi04}
{Napiwotzki} R.,  {Karl} C.~A.,  {Lisker} T.,  {Heber} U.,  {Christlieb} N.,
  {Reimers} D.,  {Nelemans} G.,   {Homeier} D.,  2004, \mn@doi [\apss]
  {10.1023/B:ASTR.0000044362.07416.6c}, \href
  {https://ui.adsabs.harvard.edu/abs/2004Ap&SS.291..321N} {291, 321}

\bibitem[\protect\citeauthoryear{{Naslim}, {Jeffery}, {Hibbert}  \&
  {Behara}}{{Naslim} et~al.}{2013}]{Naslim13}
{Naslim} N.,  {Jeffery} C.~S.,  {Hibbert} A.,   {Behara} N.~T.,  2013, \mn@doi
  [\mnras] {10.1093/mnras/stt1091}, \href
  {https://ui.adsabs.harvard.edu/abs/2013MNRAS.434.1920N} {434, 1920}

\bibitem[\protect\citeauthoryear{{Naslim}, {Jeffery}  \& {Woolf}}{{Naslim}
  et~al.}{2020}]{Naslim20}
{Naslim} N.,  {Jeffery} C.~S.,   {Woolf} V.~M.,  2020, \mn@doi [\mnras]
  {10.1093/mnras/stz3055}, \href
  {https://ui.adsabs.harvard.edu/abs/2020MNRAS.491..874N} {491, 874}

\bibitem[\protect\citeauthoryear{{O'Toole} \& {Heber}}{{O'Toole} \&
  {Heber}}{2006}]{OToole2006}
{O'Toole} S.~J.,  {Heber} U.,  2006, \mn@doi [\aap]
  {10.1051/0004-6361:20053948}, \href
  {https://ui.adsabs.harvard.edu/abs/2006A&A...452..579O} {452, 579}

\bibitem[\protect\citeauthoryear{{Oreiro}, {Ulla}, {P{\'e}rez Hern{\'a}ndez},
  {{\O}stensen}, {Rodr{\'\i}guez L{\'o}pez}  \& {MacDonald}}{{Oreiro}
  et~al.}{2004}]{Oreiro2004}
{Oreiro} R.,  {Ulla} A.,  {P{\'e}rez Hern{\'a}ndez} F.,  {{\O}stensen} R.,
  {Rodr{\'\i}guez L{\'o}pez} C.,   {MacDonald} J.,  2004, \mn@doi [\aap]
  {10.1051/0004-6361:20035844}, \href
  {https://ui.adsabs.harvard.edu/abs/2004A&A...418..243O} {418, 243}

\bibitem[\protect\citeauthoryear{{Paczy{\'n}ski}}{{Paczy{\'n}ski}}{1971}]{Paczynski1971}
{Paczy{\'n}ski} B.,  1971, \actaa, \href
  {https://ui.adsabs.harvard.edu/abs/1971AcA....21....1P} {21, 1}

\bibitem[\protect\citeauthoryear{{Pelisoli}, {Vos}, {Geier}, {Schaffenroth}  \&
  {Baran}}{{Pelisoli} et~al.}{2020}]{Pelisoli2020}
{Pelisoli} I.,  {Vos} J.,  {Geier} S.,  {Schaffenroth} V.,   {Baran} A.~S.,
  2020, \mn@doi [\aap] {10.1051/0004-6361/202038473}, \href
  {https://ui.adsabs.harvard.edu/abs/2020A&A...642A.180P} {642, A180}

\bibitem[\protect\citeauthoryear{{Petit}, {Van Grootel}, {Bagnulo},
  {Charpinet}, {Wade}  \& {Green}}{{Petit} et~al.}{2012}]{Petit2012}
{Petit} P.,  {Van Grootel} V.,  {Bagnulo} S.,  {Charpinet} S.,  {Wade} G.~A.,
  {Green} E.~M.,  2012, in {Kilkenny} D.,  {Jeffery} C.~S.,   {Koen} C.,  eds,
  Astronomical Society of the Pacific Conference Series Vol. 452, Fifth Meeting
  on Hot Subdwarf Stars and Related Objects. p.~87 (\mn@eprint {arXiv}
  {1110.5227})

\bibitem[\protect\citeauthoryear{{Przybilla}, {Nieva}  \&
  {Edelmann}}{{Przybilla} et~al.}{2006}]{Przybilla2006}
{Przybilla} N.,  {Nieva} M.~F.,   {Edelmann} H.,  2006, Baltic Astronomy, \href
  {https://ui.adsabs.harvard.edu/abs/2006BaltA..15..107P} {15, 107}

\bibitem[\protect\citeauthoryear{{Ramspeck}, {Heber}  \& {Edelmann}}{{Ramspeck}
  et~al.}{2001}]{Ramspeck2001}
{Ramspeck} M.,  {Heber} U.,   {Edelmann} H.,  2001, \mn@doi [\aap]
  {10.1051/0004-6361:20011334}, \href
  {https://ui.adsabs.harvard.edu/abs/2001A&A...379..235R} {379, 235}

\bibitem[\protect\citeauthoryear{{Randall}, {Bagnulo}, {Ziegerer}, {Geier}  \&
  {Fontaine}}{{Randall} et~al.}{2015}]{Randall2015}
{Randall} S.~K.,  {Bagnulo} S.,  {Ziegerer} E.,  {Geier} S.,   {Fontaine} G.,
  2015, \mn@doi [\aap] {10.1051/0004-6361/201425251}, \href
  {https://ui.adsabs.harvard.edu/abs/2015A&A...576A..65R} {576, A65}

\bibitem[\protect\citeauthoryear{{Rauch}}{{Rauch}}{1993}]{Rauch1993}
{Rauch} T.,  1993, \aap, \href
  {https://ui.adsabs.harvard.edu/abs/1993A&A...276..171R} {276, 171}

\bibitem[\protect\citeauthoryear{{Rauch}, {Werner}  \& {Kruk}}{{Rauch}
  et~al.}{2010}]{Rauch2010}
{Rauch} T.,  {Werner} K.,   {Kruk} J.~W.,  2010, \mn@doi [\apss]
  {10.1007/s10509-010-0304-3}, \href
  {https://ui.adsabs.harvard.edu/abs/2010Ap&SS.329..133R} {329, 133}

\bibitem[\protect\citeauthoryear{{Reed}, {Foster}, {Telting}, {{\O}stensen},
  {Farris}, {Oreiro}  \& {Baran}}{{Reed} et~al.}{2014}]{Reed2014}
{Reed} M.~D.,  {Foster} H.,  {Telting} J.~H.,  {{\O}stensen} R.~H.,  {Farris}
  L.~H.,  {Oreiro} R.,   {Baran} A.~S.,  2014, \mn@doi [\mnras]
  {10.1093/mnras/stu412}, \href
  {https://ui.adsabs.harvard.edu/abs/2014MNRAS.440.3809R} {440, 3809}

\bibitem[\protect\citeauthoryear{{Reed} et~al.,}{{Reed}
  et~al.}{2018}]{Reed2018}
{Reed} M.~D.,  et~al., 2018, \mn@doi [Open Astronomy]
  {10.1515/astro-2018-0015}, \href
  {https://ui.adsabs.harvard.edu/abs/2018OAst...27..157R} {27, 157}

\bibitem[\protect\citeauthoryear{{Reindl} et~al.,}{{Reindl}
  et~al.}{2019}]{Reindl2019}
{Reindl} N.,  et~al., 2019, \mn@doi [\mnras] {10.1093/mnrasl/sly191}, \href
  {https://ui.adsabs.harvard.edu/abs/2019MNRAS.482L..93R} {482, L93}

\bibitem[\protect\citeauthoryear{{Ricker} et~al.,}{{Ricker}
  et~al.}{2015}]{tess}
{Ricker} G.~R.,  et~al., 2015, \mn@doi [Journal of Astronomical Telescopes,
  Instruments, and Systems] {10.1117/1.JATIS.1.1.014003}, \href
  {https://ui.adsabs.harvard.edu/abs/2015JATIS...1a4003R} {1, 014003}

\bibitem[\protect\citeauthoryear{{Riello} et~al.,}{{Riello}
  et~al.}{2021}]{Riello2021}
{Riello} M.,  et~al., 2021, \mn@doi [\aap] {10.1051/0004-6361/202039587}, \href
  {https://ui.adsabs.harvard.edu/abs/2021A&A...649A...3R} {649, A3}

\bibitem[\protect\citeauthoryear{{Saffer}, {Bergeron}, {Koester}  \&
  {Liebert}}{{Saffer} et~al.}{1994}]{Saffer1994}
{Saffer} R.~A.,  {Bergeron} P.,  {Koester} D.,   {Liebert} J.,  1994, \mn@doi
  [\apj] {10.1086/174573}, \href
  {https://ui.adsabs.harvard.edu/abs/1994ApJ...432..351S} {432, 351}

\bibitem[\protect\citeauthoryear{{Sahoo} et~al.,}{{Sahoo}
  et~al.}{2020}]{Sahoo2020}
{Sahoo} S.~K.,  et~al., 2020, \mn@doi [\mnras] {10.1093/mnras/staa1337}, \href
  {https://ui.adsabs.harvard.edu/abs/2020MNRAS.495.2844S} {495, 2844}

\bibitem[\protect\citeauthoryear{{Savanov}, {Romaniuk}, {Semenko}  \&
  {Dmitrienko}}{{Savanov} et~al.}{2013}]{Savanov2013}
{Savanov} I.~S.,  {Romaniuk} I.~I.,  {Semenko} E.~A.,   {Dmitrienko} E.~S.,
  2013, \mn@doi [Astronomy Reports] {10.1134/S1063772913090059}, \href
  {https://ui.adsabs.harvard.edu/abs/2013ARep...57..751S} {57, 751}

\bibitem[\protect\citeauthoryear{{Schaffenroth}, {Classen}, {Nagel}, {Geier},
  {Koen}, {Heber}  \& {Edelmann}}{{Schaffenroth}
  et~al.}{2014}]{Schaffenroth2014}
{Schaffenroth} V.,  {Classen} L.,  {Nagel} K.,  {Geier} S.,  {Koen} C.,
  {Heber} U.,   {Edelmann} H.,  2014, \mn@doi [\aap]
  {10.1051/0004-6361/201424616}, \href
  {https://ui.adsabs.harvard.edu/abs/2014A&A...570A..70S} {570, A70}

\bibitem[\protect\citeauthoryear{{Schindewolf}, {N{\'e}meth}, {Heber},
  {Battich}, {Miller Bertolami}, {Irrgang}  \& {Latour}}{{Schindewolf}
  et~al.}{2018}]{Schindewolf2018}
{Schindewolf} M.,  {N{\'e}meth} P.,  {Heber} U.,  {Battich} T.,  {Miller
  Bertolami} M.~M.,  {Irrgang} A.,   {Latour} M.,  2018, \mn@doi [\aap]
  {10.1051/0004-6361/201732140}, \href
  {https://ui.adsabs.harvard.edu/abs/2018A&A...620A..36S} {620, A36}

\bibitem[\protect\citeauthoryear{{Schlafly}, {Meisner}  \& {Green}}{{Schlafly}
  et~al.}{2019}]{Schlafly2019}
{Schlafly} E.~F.,  {Meisner} A.~M.,   {Green} G.~M.,  2019, \mn@doi [\apjs]
  {10.3847/1538-4365/aafbea}, \href
  {https://ui.adsabs.harvard.edu/abs/2019ApJS..240...30S} {240, 30}

\bibitem[\protect\citeauthoryear{{Schneider}, {Podsiadlowski}, {Langer},
  {Castro}  \& {Fossati}}{{Schneider} et~al.}{2016}]{Schneider2016}
{Schneider} F.~R.~N.,  {Podsiadlowski} P.,  {Langer} N.,  {Castro} N.,
  {Fossati} L.,  2016, \mn@doi [\mnras] {10.1093/mnras/stw148}, \href
  {https://ui.adsabs.harvard.edu/abs/2016MNRAS.457.2355S} {457, 2355}

\bibitem[\protect\citeauthoryear{{Schneider}, {Irrgang}, {Heber}, {Nieva}  \&
  {Przybilla}}{{Schneider} et~al.}{2018}]{Schneider2018}
{Schneider} D.,  {Irrgang} A.,  {Heber} U.,  {Nieva} M.~F.,   {Przybilla} N.,
  2018, \mn@doi [\aap] {10.1051/0004-6361/201833182}, \href
  {https://ui.adsabs.harvard.edu/abs/2018A&A...618A..86S} {618, A86}

\bibitem[\protect\citeauthoryear{{Schneider}, {Ohlmann}, {Podsiadlowski},
  {R{\"o}pke}, {Balbus}, {Pakmor}  \& {Springel}}{{Schneider}
  et~al.}{2019}]{Schneider2019}
{Schneider} F. R.~N.,  {Ohlmann} S.~T.,  {Podsiadlowski} P.,  {R{\"o}pke}
  F.~K.,  {Balbus} S.~A.,  {Pakmor} R.,   {Springel} V.,  2019, \mn@doi [\nat]
  {10.1038/s41586-019-1621-5}, \href
  {https://ui.adsabs.harvard.edu/abs/2019Natur.574..211S} {574, 211}

\bibitem[\protect\citeauthoryear{{Schork}}{{Schork}}{2017}]{Schork2017}
{Schork} M.,  2017, Teacher's thesis, Friedrich-Alexander University
  Erlangen-N{\"u}rnberg

\bibitem[\protect\citeauthoryear{{Schwab}}{{Schwab}}{2018}]{Schwab2018}
{Schwab} J.,  2018, \mn@doi [\mnras] {10.1093/mnras/sty586}, \href
  {https://ui.adsabs.harvard.edu/abs/2018MNRAS.476.5303S} {476, 5303}

\bibitem[\protect\citeauthoryear{{Silvotti}, {Ostensen}  \&
  {Telting}}{{Silvotti} et~al.}{2020}]{Silvotti2020}
{Silvotti} R.,  {Ostensen} R.~H.,   {Telting} J.~H.,  2020, arXiv e-prints,
  \href {https://ui.adsabs.harvard.edu/abs/2020arXiv200204545S} {p.
  arXiv:2002.04545}

\bibitem[\protect\citeauthoryear{{Stello}, {Cantiello}, {Fuller}, {Huber},
  {Garc{\'\i}a}, {Bedding}, {Bildsten}  \& {Silva Aguirre}}{{Stello}
  et~al.}{2016}]{Stello2016}
{Stello} D.,  {Cantiello} M.,  {Fuller} J.,  {Huber} D.,  {Garc{\'\i}a} R.~A.,
  {Bedding} T.~R.,  {Bildsten} L.,   {Silva Aguirre} V.,  2016, \mn@doi [\nat]
  {10.1038/nature16171}, \href
  {https://ui.adsabs.harvard.edu/abs/2016Natur.529..364S} {529, 364}

\bibitem[\protect\citeauthoryear{{Stroeer}, {Heber}, {Lisker}, {Napiwotzki},
  {Dreizler}, {Christlieb}  \& {Reimers}}{{Stroeer} et~al.}{2007}]{Stroeer2007}
{Stroeer} A.,  {Heber} U.,  {Lisker} T.,  {Napiwotzki} R.,  {Dreizler} S.,
  {Christlieb} N.,   {Reimers} D.,  2007, \mn@doi [\aap]
  {10.1051/0004-6361:20065564}, \href
  {https://ui.adsabs.harvard.edu/abs/2007A&A...462..269S} {462, 269}

\bibitem[\protect\citeauthoryear{{Telting}, {Geier}, {{\O}stensen}, {Heber},
  {Glowienka}, {Nielsen}, {Oreiro}  \& {Frandsen}}{{Telting}
  et~al.}{2008}]{Telting2008}
{Telting} J.~H.,  {Geier} S.,  {{\O}stensen} R.~H.,  {Heber} U.,  {Glowienka}
  L.,  {Nielsen} T.,  {Oreiro} R.,   {Frandsen} S.,  2008, \mn@doi [\aap]
  {10.1051/0004-6361:200810759}, \href
  {https://ui.adsabs.harvard.edu/abs/2008A&A...492..815T} {492, 815}

\bibitem[\protect\citeauthoryear{{Tout}, {Wickramasinghe}, {Liebert},
  {Ferrario}  \& {Pringle}}{{Tout} et~al.}{2008}]{Tout2008}
{Tout} C.~A.,  {Wickramasinghe} D.~T.,  {Liebert} J.,  {Ferrario} L.,
  {Pringle} J.~E.,  2008, \mn@doi [\mnras] {10.1111/j.1365-2966.2008.13291.x},
  \href {https://ui.adsabs.harvard.edu/abs/2008MNRAS.387..897T} {387, 897}

\bibitem[\protect\citeauthoryear{{Uzundag} et~al.,}{{Uzundag}
  et~al.}{2021}]{Uzundag2021}
{Uzundag} M.,  et~al., 2021, \mn@doi [\aap] {10.1051/0004-6361/202140961},
  \href {https://ui.adsabs.harvard.edu/abs/2021A&A...651A.121U} {651, A121}

\bibitem[\protect\citeauthoryear{{Valyavin} \& {Fabrika}}{{Valyavin} \&
  {Fabrika}}{1999}]{Valyavin1999}
{Valyavin} G.,  {Fabrika} S.,  1999, in {Solheim} S.~E.,  {Meistas} E.~G.,
  eds,  Astronomical Society of the Pacific Conference Series Vol. 169, 11th
  European Workshop on White Dwarfs. p.~206

\bibitem[\protect\citeauthoryear{{Valyavin}, {Bagnulo}, {Fabrika},
  {Reisenegger}, {Wade}, {Han}  \& {Monin}}{{Valyavin}
  et~al.}{2006}]{Valyavin2006}
{Valyavin} G.,  {Bagnulo} S.,  {Fabrika} S.,  {Reisenegger} A.,  {Wade} G.~A.,
  {Han} I.,   {Monin} D.,  2006, \mn@doi [\apj] {10.1086/505781}, \href
  {https://ui.adsabs.harvard.edu/abs/2006ApJ...648..559V} {648, 559}

\bibitem[\protect\citeauthoryear{{Vos}, {{\O}stensen}, {N{\'e}meth}, {Green},
  {Heber}  \& {Van Winckel}}{{Vos} et~al.}{2013}]{Vos2013}
{Vos} J.,  {{\O}stensen} R.~H.,  {N{\'e}meth} P.,  {Green} E.~M.,  {Heber} U.,
   {Van Winckel} H.,  2013, \mn@doi [\aap] {10.1051/0004-6361/201322200}, \href
  {https://ui.adsabs.harvard.edu/abs/2013A&A...559A..54V} {559, A54}

\bibitem[\protect\citeauthoryear{{Vos}, {Vu{\v{c}}kovi{\'c}}, {Chen}, {Han},
  {Boudreaux}, {Barlow}, {{\O}stensen}  \& {N{\'e}meth}}{{Vos}
  et~al.}{2019}]{Vos2019}
{Vos} J.,  {Vu{\v{c}}kovi{\'c}} M.,  {Chen} X.,  {Han} Z.,  {Boudreaux} T.,
  {Barlow} B.~N.,  {{\O}stensen} R.,   {N{\'e}meth} P.,  2019, \mn@doi [\mnras]
  {10.1093/mnras/sty3017}, \href
  {https://ui.adsabs.harvard.edu/abs/2019MNRAS.482.4592V} {482, 4592}

\bibitem[\protect\citeauthoryear{{Vos} et~al.,}{{Vos} et~al.}{2021}]{Vos2021}
{Vos} J.,  et~al., 2021, \mn@doi [\aap] {10.1051/0004-6361/202140391}, \href
  {https://ui.adsabs.harvard.edu/abs/2021A&A...655A..43V} {655, A43}

\bibitem[\protect\citeauthoryear{{Werner}, {Dreizler}, {Heber}, {Rauch},
  {Wisotzki}  \& {Hagen}}{{Werner} et~al.}{1995}]{Werner1995}
{Werner} K.,  {Dreizler} S.,  {Heber} U.,  {Rauch} T.,  {Wisotzki} L.,
  {Hagen} H.~J.,  1995, \aap, \href
  {https://ui.adsabs.harvard.edu/abs/1995A&A...293L..75W} {293, L75}

\bibitem[\protect\citeauthoryear{{Werner}, {Reindl}, {Geier}  \&
  {Pritzkuleit}}{{Werner} et~al.}{2022}]{Werner2022}
{Werner} K.,  {Reindl} N.,  {Geier} S.,   {Pritzkuleit} M.,  2022, \mn@doi
  [\mnras] {10.1093/mnrasl/slac005}, \href
  {https://ui.adsabs.harvard.edu/abs/2022MNRAS.511L..66W} {511, L66}

\bibitem[\protect\citeauthoryear{{Woltjer}}{{Woltjer}}{1964}]{Woltjer1964}
{Woltjer} L.,  1964, \mn@doi [\apj] {10.1086/148028}, \href
  {https://ui.adsabs.harvard.edu/abs/1964ApJ...140.1309W} {140, 1309}

\bibitem[\protect\citeauthoryear{{Wurster}, {Bate}  \& {Price}}{{Wurster}
  et~al.}{2018}]{Wurster2018}
{Wurster} J.,  {Bate} M.~R.,   {Price} D.~J.,  2018, \mn@doi [\mnras]
  {10.1093/mnras/sty2438}, \href
  {https://ui.adsabs.harvard.edu/abs/2018MNRAS.481.2450W} {481, 2450}

\bibitem[\protect\citeauthoryear{{Yu}, {Zhang}  \& {L{\"u}}}{{Yu}
  et~al.}{2021}]{Yu2021}
{Yu} J.,  {Zhang} X.,   {L{\"u}} G.,  2021, \mn@doi [\mnras]
  {10.1093/mnras/stab1063}, \href
  {https://ui.adsabs.harvard.edu/abs/2021MNRAS.504.2670Y} {504, 2670}

\bibitem[\protect\citeauthoryear{{Zhang} \& {Jeffery}}{{Zhang} \&
  {Jeffery}}{2012}]{Zhang2012}
{Zhang} X.,  {Jeffery} C.~S.,  2012, \mn@doi [\mnras]
  {10.1111/j.1365-2966.2011.19711.x}, \href
  {https://ui.adsabs.harvard.edu/abs/2012MNRAS.419..452Z} {419, 452}

\makeatother
\end{thebibliography}



\appendix

\section{Multi-component spectroscopic fits}

\begin{table*}
\caption{Atmospheric parameters from fits with one, two, or three components to the co-added WHT/ISIS spectra of the three targets.}
\label{tab:components}
\begin{center}
\begin{tabular}{l c c c c c c c c c} 
\hline
Star & \teff/K & \logg\ & \logy & $B_1$/kG & $B_2$/kG  & $B_3$/kG & $A_2/A_1$ & $A_3/A_1$ & $\chi ^2 _r$\\
\hline 
\jofour    & 46730 & 6.02 & $-$0.15 & 280 & -- & -- & -- & -- & 2.56 \\
\jofour    & 46460 & 5.95 & $-$0.12 & 266 & 420 & -- & 0.24 & -- & 2.31 \\
\jofour    & 46430 & 5.96 & $-$0.13 & 262 & 377 & 469 & 0.21 & 0.10 & 2.45  \\
\hline 
\jthirteen & 48880 & 6.07 & +0.22 & 415 & -- & -- & -- & -- & 2.66 \\
\jthirteen & 47920 & 5.87 & +0.32 & 384 & 571 & -- & 0.56 & -- & 1.99 \\
\jthirteen & 47790 & 5.84 & +0.33 & 364 & 584 & 442 & 0.67 & 0.61 & 1.89 \\
\hline 
\jsixteen  & 46620 & 6.08 & +0.06 & 340 & -- & -- & -- & -- & 2.33 \\
\jsixteen  & 45980 & 6.03 & +0.05 & 291 & 395 & -- & 0.82 & -- & 2.10 \\
\jsixteen  & 45700 & 5.95 & +0.06 & 284 & 377 & 523 & 0.90 & 0.27 & 2.07 \\
\hline
\end{tabular}
\end{center}
\end{table*}

\begin{table}
\caption{The magnetic fields of the individual components and their relative surface ratio for each of the three stars in our best-fit model to the individual WHT/ISIS spectra.
The uncertainties for the surface ratios are one sigma statistical, whereas the uncertainties on the magnetic field strengths are estimated systematic uncertainties. }
\label{tab:components2}
\begin{center}
\begin{tabular}{l r r r}
\hline
{} & \jofour & \jthirteen & \jsixteen   \\
\hline
$B_1$ (kG) & $270 \pm 15$  & $370 \pm 20$ & $292 \pm 15$ \\
$B_2$ (kG) & $430 \pm 30$   & $581 \pm 20$ & $390 \pm 15$ \\
$B_3$ (kG) & --   & $439 \pm 20$ & -- \\[2pt]
$A_2/A_1$  & $0.260^{+0.014}_{-0.014}$   & $0.70^{+0.13}_{-0.05}$ & $0.81^{+0.16}_{-0.08}$ \\[2pt]
$A_3/A_1$  & --   & $0.56^{+0.23}_{-0.08}$ & -- \\
\hline
\end{tabular}
\end{center}
\end{table}

\section{SED fits}
\label{sec:phot}

\clearpage

\begin{figure}
	\includegraphics[width=\columnwidth]{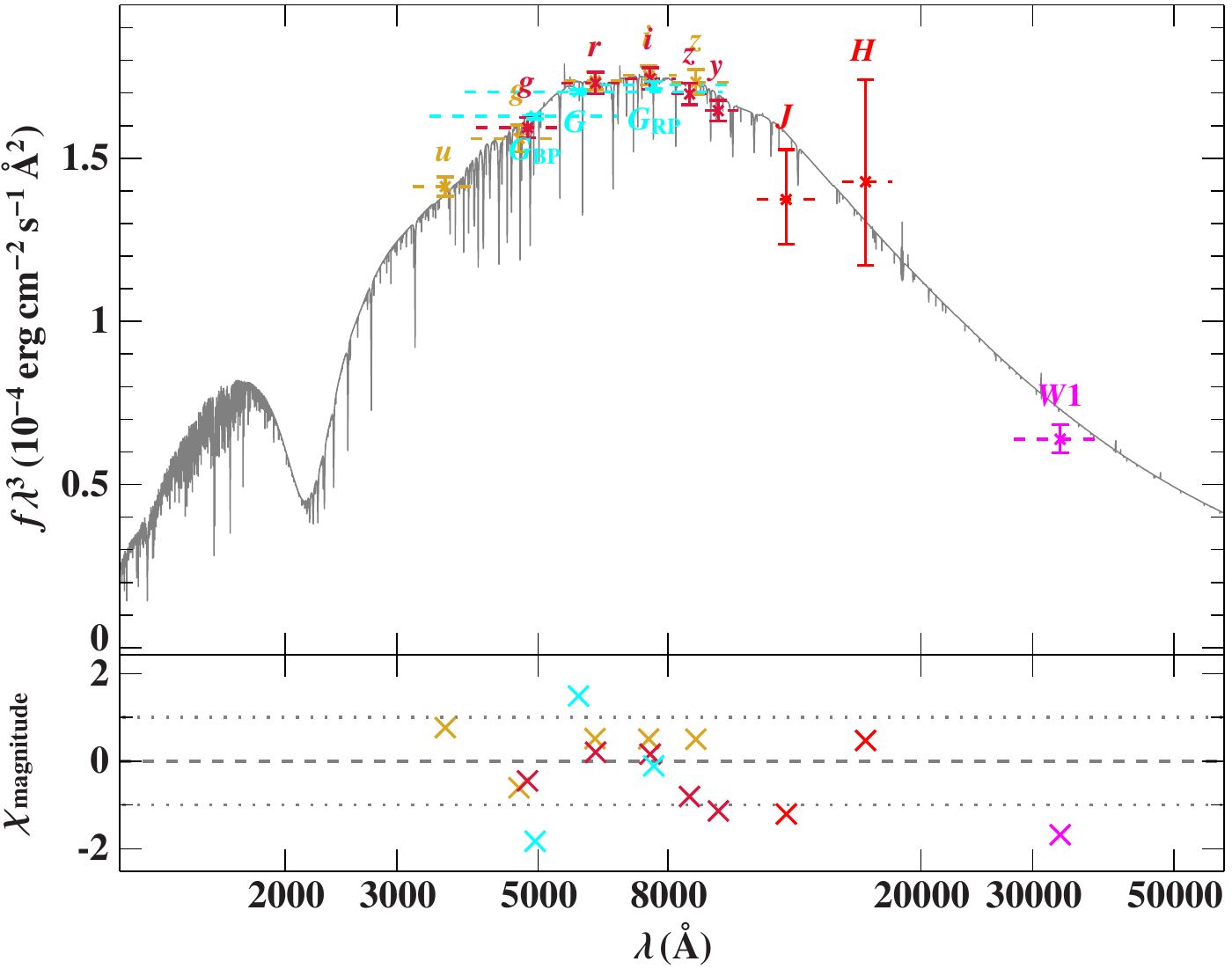}
    \caption{SED fit for \jofour. The grey line shows the best-fit, while filter-averaged flux measurements are shown by dashed horizontal lines. Residuals are shown in the bottom panel. The photometric systems are colour-coded: SDSS \citep[ochre,][]{Alam2015}, Pan-STARRS \citep[dark red,][]{Magnier2020}, {\it Gaia} EDR3 \citep[cyan,][]{Riello2021}, 2MASS \citep[red,][]{Cutri2003}, and WISE \citep[magenta,][]{Schlafly2019}.}
    \label{fig:sed04}
\end{figure}

\begin{figure}
	\includegraphics[width=\columnwidth]{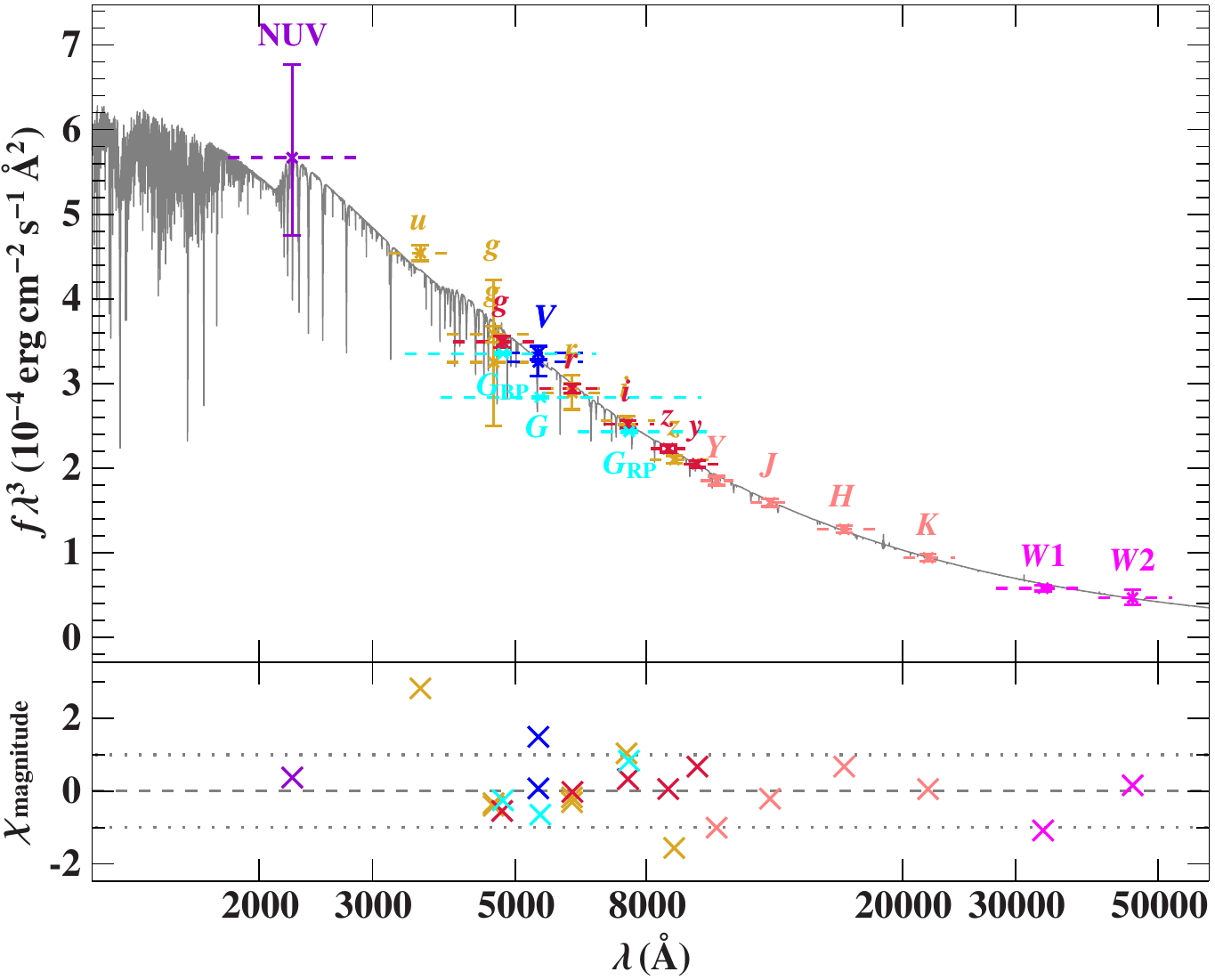}
    \caption{SED fit for \jthirteen. Like for Fig.~\ref{fig:sed04}, we show the model in grey and the filter-averaged flux measurements as dashed lines. The photometric systems are GALEX \citep[purple,][]{Bianchi2017}, SDSS \citep[ochre,][]{Alam2015,Henden2016}, Pan-STARRS \citep[dark red,][]{Magnier2020}, Johnson \citep[blue,][]{Kilkenny1988,Henden2016}, {\it Gaia} EDR3 \citep[cyan,][]{Riello2021}, UKIDSS \citep[pink,][]{ukidss}, and WISE \citep[magenta,][]{Schlafly2019}. }
    \label{fig:sed13}
\end{figure}

\begin{figure}
	\includegraphics[width=\columnwidth]{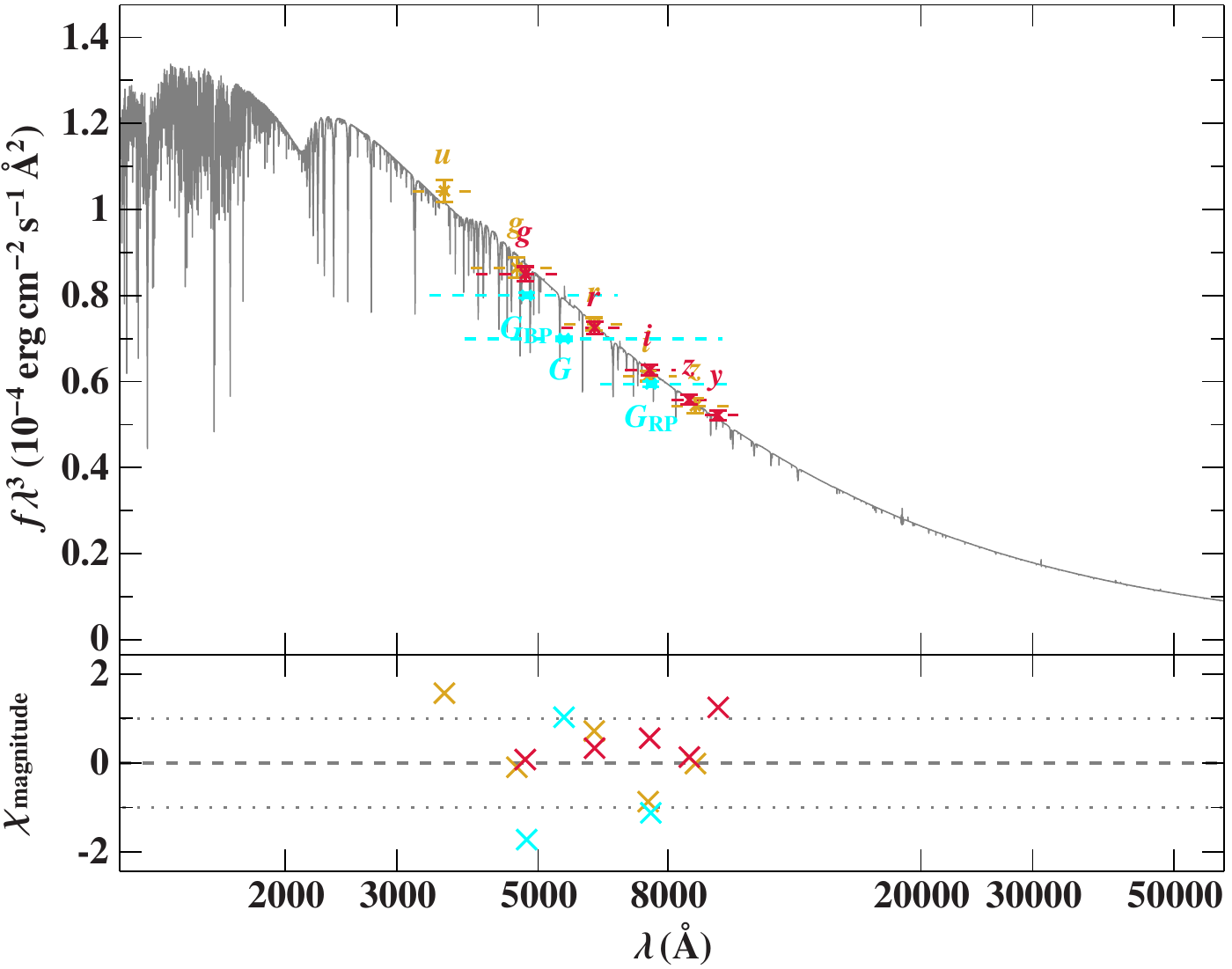}
    \caption{SED fit for \jsixteen\ using SDSS \citep[ochre,][]{Alam2015}, Pan-STARRS \citep[dark red,][]{Magnier2020}, and  {\it Gaia} EDR3 \citep[cyan,][]{Riello2021}.}
    \label{fig:sed16}
\end{figure}

\section{Additional light curves}
\label{sec:lcs}

\begin{figure*}
	\includegraphics[width=0.9\textwidth]{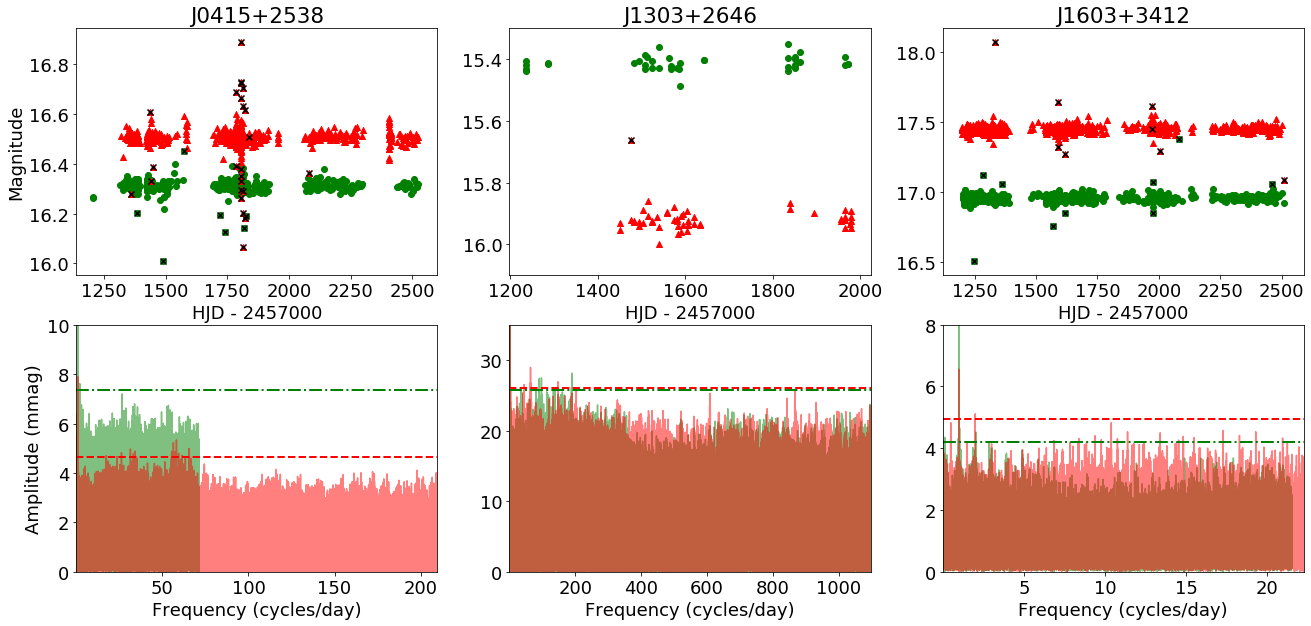}
    \caption{The light curves for the $r$ (red triangles) and $g$ (green circles) filters are shown in the top panel, with excluded datapoints marked by crosses. The bottom panels show the Fourier transform. The only peaks significantly above the detection threshold of four times the average (red dashed line for $r$, green dot-dashed line for $g$) are multiples of one-day aliases, seen clearly in particular for \jsixteen.}
    \label{fig:ztf}
\end{figure*}

\begin{figure*}
	\includegraphics[width=0.9\textwidth]{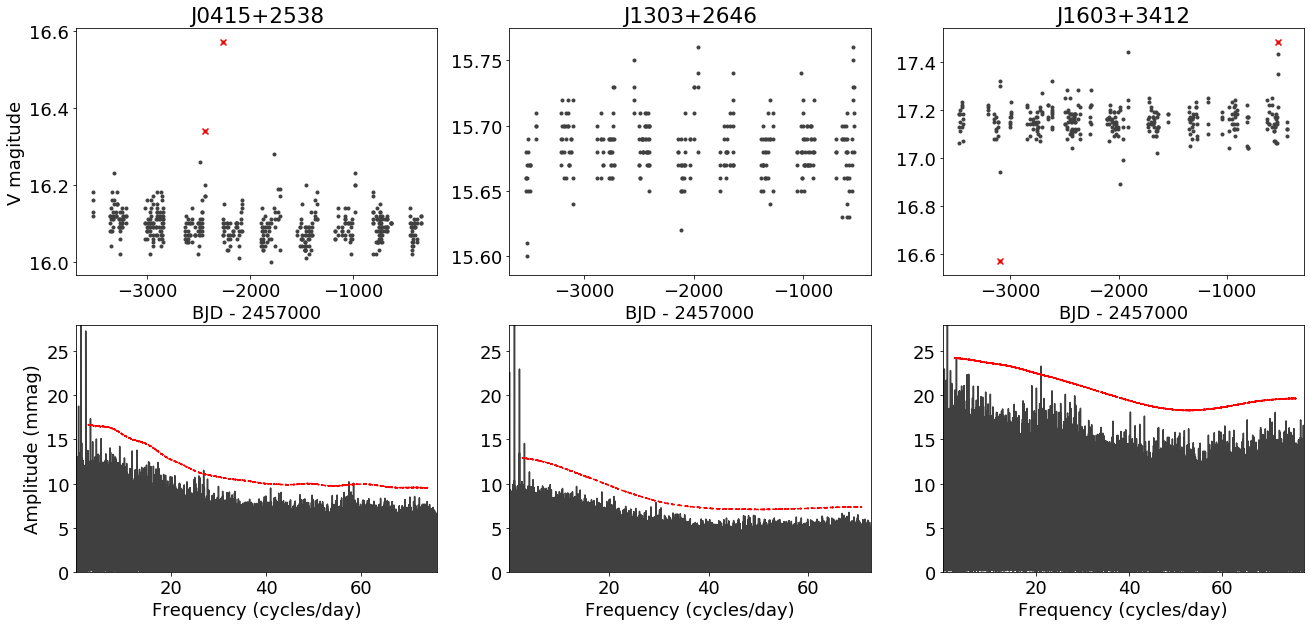}
    \caption{CRTS light curves are shown in the top panels, with the bottom panels showing the respective Fourier transforms. The dashed line indicating the threshold here was calculated as four times the average amplitude in a 5~cycles/day window, given the visible varying amplitude over the frequency spectrum. Multiples of one-day aliases are seen for all light curves. Some other marginal peaks appear slightly above the threshold, but they are not seen in the TESS or ZTF data.}
    \label{fig:crts}
\end{figure*}

\clearpage

\section{Hot subwarfs probed for magnetic fields}
\label{sec:specs}

Table~\ref{tab:litsds} lists, to the best of our knowledge, all hot subdwarfs with determined atmospheric parameters that have upper limits or disputed claims of a magnetic field from spectropolarimetry. In addition, they all have spectra of similar quality or better than the stars discussed here, which would reveal Zeeman splitting for fields $\sim 50$~kG or more. Among the hot subdwarfs, the sdB HD~76431 has been studied by spectropolarimitry most extensively \citep{Elkin1998, Petit2012, Landstreet2012, Chountonov2012} at many epochs, but no detection of a significant magnetic field was reported. \citet{Chountonov2012} estimated the detection limit at 100--200~G.
For other stars in Table~\ref{tab:litsds}, no field could be reported at upper detection limits of 1~kG or better (see also Sect. \ref{sec:magnetic}). For the four sdBs studied by \citet{Kawka2007} the limits turned out to be somewhat higher at several kG. 
The distribution of the stars listed in Table~\ref{tab:litsds} in the Kiel diagram is shown in Fig. \ref{fig:kiel_magnetic}. All subtypes are represented (sdB, sdOB, sdO, He-sdB, as well as both variants of He-sdO, that is iHe and eHe-sdOs), though the majority are sdBs. Also some more luminous subdwarfs (e.g. LS~IV-12~1, LSE~263, and LSE~153, marked with the prefix ``l'') are included which probably evolved from the AGB. HD~188112 is an underluminous sdB of too low mass for core helium burning to ignite, and Balloon~09010~0001 is a large amplitude pulsating (V361Hya) star \citep{Telting2008}. 
The main types of binaries are also all represented (white dwarf or low-mass companion with short orbital period, main sequence or giant companions in long orbital period systems), with only seven stars lacking sufficient \vrad\ measurements to allow conclusive remarks about binary status. An unconfirmed detection of a variable magnetic field was reported for BD+75 325 (see Sect. \ref{sec:magnetic}). 

\begin{table*}
\caption{Hot subdwarfs with well-determined atmospheric parameters and upper limits on magnetic fields, typically of the order of a few kG. The \vrad\ variability is inferred from multi-epoch observations indicated in the notes. The orbital period is given in days when determined, and the entry ``no'' indicates no \vrad\ variations detected on long time scales ($>$months).}
\label{tab:litsds}
\begin{center}
\begin{tabular}{l c c c c c c c c c}
\hline
Name  & Spectral  & \vrad\  & \teff & \logg & \logy & \multicolumn{2}{c}{References} \\
      &     class     & variablity                 &       &       &       & Atmospheric parameters & $B$ limit \\
\hline
BD+75~325             &    iHe-sdO      &     no$^{\mathrm{s17}}$    & $52000\pm2000$ & $5.50\pm0.20$ & +0.00          & \citet{Lanz1997} & \citet{Elkin1996,Elkin1998} \\
HD~128220             &    lHe-sdO+GIII      &  871.78$^{\mathrm{HH}}$       & $40600\pm400$  & $4.5\pm0.1$   & $0.30\pm0.05$  & \citet{Rauch1993} & \citet{Elkin1998} \\
BD+25~4655            &  eHe-sdO      &   no$^{\mathrm{E}}$    &       $39500\pm1000$         &  $5.8\pm0.1$  & $1.55\pm0.15$  &  \citet{Dorsch2022b}  & \citet{Elkin1998} \\
Feige~87              &  sdB+G      &   936$^{\mathrm{V}}$      & $27270\pm500$ & $5.47\pm0.15$ &     $-2.56^{+0.22}_{-0.50}$           & \citet{Vos2013}     & \citet{Elkin1998} \\
HD~76431               &    sdB      &     no$^{\mathrm{R,Kh,CG}}$ & $31180\pm220$  & $4.67\pm0.03$ & $-1.58\pm0.05$ & \citet{Khalack2014} & \citet{Chountonov2012} \\
GD~687                 &    sdB+WD   &     0.37765$^{\mathrm{G}}$     & $24350\pm360$  & $5.32\pm0.05$ & $-2.38 $       & \citet{Lisker2005} & \citet{Kawka2007} \\
GD~1669                &    sdB      &       no$^{\mathrm{GH}}$                     & $34126\pm360$  & $5.77\pm0.05$ & $-1.36$        & \citet{Lisker2005} & \citet{Kawka2007} \\
GD~108                 &    sdB+?    &     3.18095$^{\mathrm{C}}$     & $27760\pm670$  & $5.60\pm0.11$ & $<-3.0$        & \citet{Kawka2007} & \citet{Kawka2007} \\
WD~1153-484            &    sdB      &                            & $30080\pm660$  & $5.15\pm0.10$ & $<-3.0$        & \citet{Kawka2007} & \citet{Kawka2007} \\
SB~290                 &    sdB+K    &   uncertain$^{\mathrm{G}}$ & $26300\pm100$  & $5.31\pm0.01$ & $-2.52\pm0.08$ & \citet{Geier2013} & \citet{Landstreet2012} \\
HD~4539               &    sdB      &     no$^{\mathrm{S,K,E}}$  & $23200\pm100$  & $5.20\pm0.01$ & $-2.27\pm0.24$ & \citet{Schneider2018} & \citet{Landstreet2012} \\
PHL~932               &    sdB      &     no$^{\mathrm{K,E}}$    & $33644\pm500$  & $5.74\pm0.05$ & $-1.64\pm0.05$ & \citet{Lisker2005} & \citet{Landstreet2012} \\
PG~0133+114            &    sdB+WD   &     1.23787$^{\mathrm{E}}$     & $30073\pm201$  & $5.70\pm0.04$ & $-2.14\pm0.04$ & \citet{Luo2021} & \citet{Landstreet2012} \\
SB~707                 &    sdB+WD   &     5.85$^{\mathrm{E}}$     & $35400\pm500$  & $5.90\pm0.05$ & $-2.90\pm0.10$ & \citet{OToole2006} & \citet{Landstreet2012} \\
PG~0342+026            &    sdB      &     no$^{\mathrm{E,S}}$    & $26000\pm1100$ & $5.59\pm0.12$ & $-2.69\pm0.10$ & \citet{Geier2013} & \citet{Landstreet2012} \\
HD~127493              &    iHe-sdO   &     no$^{\mathrm{E}}$      & $42070\pm180$  & $5.61\pm0.04$ & $ 0.33\pm0.06$ & \citet{Dorsch2019} & \citet{Landstreet2012} \\
HD~149382              &    sdB      &     no$^{\mathrm{J}}$      & $34200\pm1000$ & $5.89\pm0.15$ & $-1.60\pm0.10$ & \citet{Saffer1994} & \citet{Landstreet2012} \\
HD~171858              &    sdB+WD   &     1.63280$^{\mathrm{E}}$     & $27200\pm800$  & $5.30\pm0.10$ & $-2.84\pm0.1 $ & \citet{Geier2010} & \citet{Landstreet2012} \\
HD~188112              &    sdB+WD   &     0.6065812$^{\mathrm{E}}$     & $21500\pm500$  & $5.66\pm0.06$ & $-5.00       $ & \citet{Heber2003} & \citet{Landstreet2012} \\
HD~205805              &    sdB      &     no$^{\mathrm{E}}$      & $25000\pm500$  & $5.00\pm0.10$ & $-2.00\pm0.2 $ & \citet{Przybilla2006} & \citet{Landstreet2012} \\
JL~87                  &    iHe-sdB   &     no$^{\mathrm{E}}$      & $25800\pm1000$ & $4.80\pm0.30$ & $ 0.33       $ & \citet{Ahmad2007} & \citet{Landstreet2012} \\
$[$CW 83$]$~0512-08   &    sdB     &     no$^{\mathrm{E,S}}$    & $38400\pm1100$ & $5.77\pm0.12$ & $-0.73\pm0.10$ & \citet{Geier2013} & \citet{Landstreet2012} \\
CPD-64 481            &    sdB+BD?  &     0.27726315$^{\mathrm{Sch}}$     & $27500\pm500$  & $5.60\pm0.05$ & $-2.50\pm0.10$ & \citet{OToole2006} & \citet{Landstreet2012} \\
CD-31 4800            &    eHe-sdO   &     no$^{\mathrm{E}}$      & $42230\pm300$  & $5.60\pm0.1 $ & $ 2.61\pm0.20$ & \citet{Schindewolf2018} & \citet{Landstreet2012} \\
PG~0909+276            &    sdOB     &     no$^{\mathrm{E}}$      & $35500\pm500$  & $6.09\pm0.05$ & $-1.00\pm0.10$ & \citet{Geier2013} & \citet{Landstreet2012} \\
LS~IV-12~1             &    lsdO      &     no$^{\mathrm{E}}$      & $60000\pm5000$ & $4.50\pm0.50$ & $-0.95\pm0.20$ & \citet{Heber1987} & \citet{Landstreet2012} \\
LSE~263                &    lHe-sdO   &     no$^{\mathrm{K}}$      & $70000\pm2500$ & $4.90\pm0.25$ &  $>$+1.0 & \citet{Husfeld1989} & \citet{Landstreet2012} \\
LSE~153                &    lHe-sdO   &                            & $70000\pm1500$ & $4.75\pm0.15$ &      $>$+1.0          & \citet{Husfeld1989} & \citet{Landstreet2012} \\
BD+28~4211            &    sdO      &     no$^{\mathrm{L,H}}$    & $81300\pm1200$ & $6.52\pm0.05$ & $-1.12\pm0.05$ & \citet{Latour2015} & \citet{Landstreet2012} \\
EC~11481-2303         &    sdO      &                            & $55000\pm5000$ & $5.8\pm0.3$ & $-2.0\pm0.3$   & \citet{Rauch2010} & \citet{Landstreet2012} \\
SB~410                 &    sdB+WD   &     0.8227$^{\mathrm{E}}$     & $27600\pm500$  & $5.43\pm0.05$ & $-2.71\pm0.10$ & \citet{Geier2010} & \citet{Mathys2012} \\
SB~459                 &    sdB      &                            & $24900\pm500$  & $5.35\pm0.10$ & $-2.58\pm0.10$ & \citet{Sahoo2020} & \citet{Mathys2012} \\
LB~1516                &    sdB+WD   &     10.3598$^{\mathrm{G2}}$     & $25200\pm1100$ & $5.41\pm0.12$ & $-2.78\pm0.10$ & \citet{Geier2013} & \citet{Mathys2012} \\
JL~194                 &    sdB      &     no$^{\mathrm{E}}$      & $25770\pm380$  & $5.21\pm0.06$ & $-2.69\pm0.06$ & \citet{Uzundag2021} & \citet{Mathys2012} \\
GD~1110                &    sdB+dM/BD&     0.3131$^{\mathrm{Sch}}$     & $26500\pm1100$ & $5.38\pm0.12$ & $-2.54\pm0.10$ & \citet{Geier2013} & \citet{Mathys2012} \\
SB~815                &    sdB      &     no$^{\mathrm{K}}$      & $27200\pm550$  & $5.39\pm0.10$ & $-2.94\pm0.01$ & \citet{Schneider2018} & \citet{Mathys2012} \\
Feige~66              & sdB      &                            & $33220\pm370$  & $6.14\pm0.08$ & $-1.61\pm0.11$ & \citet{Lei2018} & \citet{Petit2012} \\
LS~IV-14~116           &    iHe-sdO   &     no$^{\mathrm{JS,Ra}}$  & $35500\pm1000$ & $5.85\pm0.10$ & $-0.60\pm0.10$ & \citet{Dorsch2020} & \citet{Randall2015} \\
Balloon~09010~0001     &    sdB      &     0.0041$^{\mathrm{T}}$      & $29446\pm500$  & $5.33\pm0.1$ & $-2.54\pm0.2$  & \citet{Oreiro2004} & \citet{Savanov2013} \\
Feige~34              &    sdO      &                            & $62550\pm600$  & $5.99\pm0.03$ & $-1.79\pm0.04$ & \citet{Latour2018} & \citet{Valyavin2006} \\
\hline
\end{tabular}
\end{center}
{\bf Notes:} 
E = \citet{Edelmann2005} (variables published, non-variables: priv. com.),
S17 =\citet{Schork2017},
S = \citet{Silvotti2020},
J = \citet{Jacobs2011},
K = \citet{Kawka2015},
Kh = \citet{Khalack2014},
R = \citet{Ramspeck2001},
Ra = \citet{Randall2015},
L = \citet{Latour2015},
H = \citet{Herbig1999},
JS = \citet{Jeffery2015},
R = \citet{Randall2015},
C = \citet{Copperwheat2011},
G = \citet{Geier2010spy},
\uh{GH = \citet{Geier2012}},
T = \citet{Telting2008},
HH=\citet{Howarth1990},
CG=\citet{Chountonov2012},
G2=\citet{Geier2014},
Sch=\citet{Schaffenroth2014}
V=\citet{Vos2013}.
\end{table*}



\bsp	
\label{lastpage}
\end{document}